\documentclass[12pt,a4paper]{article}
\usepackage[utf8]{inputenc}
\usepackage[english]{babel}
\usepackage{amsmath}
\usepackage{amsfonts}
\usepackage{amssymb}
\usepackage[pdftex]{graphicx}
\usepackage[font=small,labelfont=bf]{caption}
\usepackage{epstopdf}
\usepackage[T1]{fontenc}
\usepackage{etoolbox}
\usepackage{cleveref}
\usepackage{multirow}
\usepackage{amsbsy}
\usepackage[top=1.5in, bottom=1.5in, left=0.8in, right=0.8in]{geometry}
\usepackage{color}
\usepackage[pagewise]{lineno}
\usepackage{cleveref}
\usepackage{subfigure}
\usepackage{url}
\usepackage{authblk}

\allowdisplaybreaks

\DeclareUnicodeCharacter{00A0}{~}
\DeclareUnicodeCharacter{2212}{~}

\newcommand{\ud}{\mathrm{d}}
\newcommand{\ue}{\mathrm{e}}

\newcommand{\be}{\begin{equation}}
\newcommand{\ee}{\end{equation}}
\newcommand{\bea}{\begin{eqnarray}}
\newcommand{\eea}{\end{eqnarray}}
\newcommand{\beas}{\begin{eqnarray*}}
\newcommand{\eeas}{\end{eqnarray*}}
\newcommand{\bi}{\begin{itemize}}
\newcommand{\ei}{\end{itemize}}
\newcommand{\lb}{\label}
\newcommand\nc{\newcommand}
\nc{\Oc}{\mathcal{O}}
\nc{\Omo}{\Omega_{\tiny{\mbox{out}}}}
\nc{\Omi}{\Omega_{\tiny{\mbox{in}}}}
\nc{\Omit}{\Omega_{\tiny{\mbox{int}}}}
\nc{\Pio}{\Pi_{\tiny{\mbox{out}}}}
\nc{\Pii}{\Pi_{\tiny{\mbox{in}}}}
\nc{\Piit}{\Pi_{\tiny{\mbox{int}}}}
\nc\pa{\partial}
\nc\pad[2]{\frac{\pa #1}{\pa #2}}
\nc\padd[2]{\frac{\pa^2 #1}{\pa{#2}^2}}
\nc\nd[2]{\frac{\mathrm{d} #1}{\mathrm{d} #2}}
\nc\ndd[2]{\frac{d^2 #1}{d {#2}^2}}
\nc\pat[2]{\frac{D #1}{D
#2}}
\nc\ov{\overline}
\nc\degree{^{\circ}} \nc\ord[1]{{\cal
O}(#1)} \nc\ra{\rightarrow} \nc\Ra{\Rightarrow} \nc\dint{{\mbox ~
d}}
\nc\dg{{\dot \gamma}}

\newcommand{\etal}{\emph{et al.}$\;$}

\newenvironment{subeq}[1]{
\begin{subequations}
#1
\end{subequations}
}

\graphicspath{{results/figs/}}

\newcommand*\patchAmsMathEnvironmentForLineno[1]{%
  \expandafter\let\csname old#1\expandafter\endcsname\csname #1\endcsname
  \expandafter\let\csname oldend#1\expandafter\endcsname\csname end#1\endcsname
  \renewenvironment{#1}%
     {\linenomath\csname old#1\endcsname}%
     {\csname oldend#1\endcsname\endlinenomath}}%
\newcommand*\patchBothAmsMathEnvironmentsForLineno[1]{%
  \patchAmsMathEnvironmentForLineno{#1}%
  \patchAmsMathEnvironmentForLineno{#1*}}%
\AtBeginDocument{%
\patchBothAmsMathEnvironmentsForLineno{equation}%
\patchBothAmsMathEnvironmentsForLineno{align}%
\patchBothAmsMathEnvironmentsForLineno{flalign}%
\patchBothAmsMathEnvironmentsForLineno{alignat}%
\patchBothAmsMathEnvironmentsForLineno{gather}%
\patchBothAmsMathEnvironmentsForLineno{multline}%
}

\begin{document}

\title{The Stefan problem with variable thermophysical properties and phase change temperature}

\renewcommand*{\Affilfont}{\small\itshape}
\author[1,2]{T.~G.~Myers\thanks{tmyers@crm.cat}}
\author[1,3]{M.~G.~Hennessy}
\author[1,2]{M.~Calvo-Schwarzw\"alder}

\affil[1]{Centre de Recerca Matem\`atica, Campus de Bellaterra, Edifici C, 08193 Bellaterra, Spain}
\affil[2]{Department de Matem\`atica Aplicada, Universitat Polit\`ecnica de Catalunya,  08028 Barcelona, Spain}
\affil[3]{Mathematical Institute, University of Oxford, Andrew Wiles Building, Radcliffe Observatory Quarter, Woodstock Road, Oxford, OX2 6GG, United Kingdom}

\date{\today}

\maketitle

\begin{abstract}

In this paper we formulate a Stefan problem appropriate when the thermophysical properties are distinct in each phase and the phase-change temperature is size or velocity dependent. Thermophysical properties invariably take different values in different material phases but this is often ignored for mathematical simplicity. Size and velocity dependent phase change temperatures are often found at very short length scales, such as nanoparticle melting or dendrite formation; velocity dependence occurs in the solidification of supercooled melts. To illustrate the method we show how the governing equations may be applied to a standard one-dimensional problem and also the melting of a spherically symmetric nanoparticle. Errors which have propagated through the literature are  highlighted. By writing the system in non-dimensional form we are able to study the large Stefan number formulation and an energy-conserving one-phase reduction. The results from the various simplifications and assumptions are compared with those from a finite difference numerical scheme. Finally, we briefly discuss the failure of Fourier's law at very small length and time-scales and provide an alternative formulation which takes into account the finite time of travel of heat carriers (phonons) and the mean free distance between collisions.

\end{abstract}






\section{Introduction}

The classical Stefan problem is described by two heat equations, one in each of the two material phases. The domain over which each equation holds is determined by an energy balance, the Stefan condition, which equates the energy released during the phase change with that conducted away through either phase. Thermophysical properties are assumed constant (often taking the same value in each phase) and the melt temperature is also constant. Whilst exact solutions do exist for certain configurations and boundary conditions, there is possibly just one physically useful one: the Neumann solution. This involves a semi-infinite material with constant thermophysical properties changing phase due to an instantaneous switch in the  boundary temperature, which subsequently stays at the same, constant value. The classic problem, and Neumann solution, are described in many texts, see \cite{Alex,Davis2001,Hill1987} for example.

In practice an instantaneous switch to a new temperature is impossible to achieve and it is difficult to maintain a constant boundary temperature throughout a phase-change process. Thermophysical properties are temperature dependent (admittedly sometimes only weakly so), but they always differ between phases (often quite significantly). There also exist situations where properties are size or velocity dependent. This may occur, for example, when the curvature is high, such as in dendritic growth or nanoscale phase change, or with the solidification of supercooled materials.

The size dependence of material properties at small length-scales has been demonstrated experimentally, by molecular dynamics simulations, and also theoretically. One of the main reasons for this  dependence is the large ratio of surface to interior atoms. Interior atoms are free to bond with other atoms on all sides. Surface atoms are exposed to the environment and do not have the ability to bond on all sides, consequently they behave differently to interior atoms. At the macro-scale the very high ratio of interior to surface atoms means that the interior dominates the material's behaviour. However, for sufficiently small samples the ratio decreases until surface properties become important. At this stage size dependence must be considered in any material description. This argument is equally valid whenever the surface curvature is sufficiently high that the local surface to bulk ratio is large.

Properties that vary with size include the catalytic, ferromagnetic, and  mechanical characteristics, colour, surface tension, phase change temperature, and latent heat \cite{Guisbiers2012,Myers16}. For example, colloidal gold with particles  smaller than 40 nm has a colour ranging from clear to yellow, between 40--100 nm it has a reddish hue, between 100--120 nm  the colour becomes purple. In terms of phase change, experiments have shown
a decrease in the melt temperature of approximately 500 K below the bulk value for gold nanoparticles with radii of the order of 1 nm \cite{Buffat1976} and 70 K and 200 K in tin and lead nanoparticles, respectively \cite{David1995}.  Molecular dynamics simulations have shown  a decrease of 800 K (approximately 60\%) for gold nanoparticles of radius 0.8 nm \cite{Shim2002}. It has also been suggested that latent heat and surface tension decrease with size. For tin decreases in the latent heat of the order 70\% below the bulk value were reported by \cite{Lai1996} although subsequent refinements to the technique showed even greater reductions \cite{Jiang2006}. Sun and Simon \cite{Sun2007} concluded that the melt temperature is well approximated by the Gibbs--Thomson relation while the latent heat decrease is much greater than theoretically predicted. Ribera and Myers \cite{Ribera2016} review theoretical predictions for latent heat reduction and conclude that none of the models  examined match the experimental data. Decreases in surface tension are usually less significant than those observed in the latent heat and melt temperature, typically of the order  15\% below the bulk value for $R = 5$ nm \cite{Tolman1949}. With supercooling the melt temperature $T_I$ varies nonlinearly with the fluid velocity. In \cite{Font13b} examples are provided for copper, where the front velocity varies between 0 and 2.9~m/s as the degree of supercooling varies between 0 and 1000~K.

There has been particular focus on spherical nanoparticle melting, since this provides a simple framework to investigate the effect of size-dependent properties. The theoretical investigations on spherical nanoparticle phase change  \cite{Back2014b,Back2014a,Font13a,McCue2009,Wu2009,Wu2009a} have the same density in both phases but include melting point depression. The mathematical model in these papers involves standard heat equations and a Stefan condition with an \lq effective' latent heat of the form $L_m + \Delta c (T_I-T_m)$, where $T_m$ is the bulk melt temperature, $\Delta c$ is the jump in specific heat between the liquid and solid phases and $T_I(t)$ the temperature at the phase change interface. This expression is also employed in studies of the solidification of supercooled materials \cite{Alex,Myers2015}.
Melting point depression and constant (but different) densities in each phase are included in the model of \cite{Font15}. The change in density forces fluid motion, so introducing advection into the liquid heat equation and kinetic energy into the Stefan condition. The generation of kinetic energy means there is less energy  available for the phase change which results in significantly longer melt times. The model of \cite{Font15} is an extension of the Cartesian version of \cite{Alex} which, due to a lack of curvature effects, does not include melting point depression. The thesis of Back
\cite[\S 7.1-7.4]{Back2014} discusses a number of variations to the standard Stefan problem including the advection/kinetic energy terms. A model is also discussed including a size-dependent latent heat, following the formula of \cite{Lai1996}, which is released at the bulk melt temperature.
In \cite{Myers16} the governing equations include an effective latent heat which  is the sum of the size-dependent latent heat, the kinetic energy, and the energy required to make new surface. Surface energy is also included in the model of \cite{Davis2001}.
In \cite{Ribera2016} it is shown that the formula of \cite{Lai1996} underestimates the value of latent heat near the bulk value (where the data is most reliable). They also demonstrate that other formulations, such as those of \cite{Shin2014,Xiong2011}, also provide a poor  match with experimental data. They go on to propose an exponential fit to the data for tin nanoparticles which provides excellent agreement in the limit of large particles and only shows a noticeable discrepancy for $ R<8$ nm. This fit is used in the model developed in \cite{Myers16} to demonstrate significantly faster melt times than those with a constant latent heat. However,  the well-known liquid skin model \cite{Lai1996} indicates that the experimental calculation of latent heat does not coincide with the actual latent heat release. This is because the liquid skin is thought to form spontaneously, with a resultant decrease in surface energy and also an energy cost in forming the skin.  Thus, the experimentally measured latent heat should be interpreted as an effective latent heat rather than the actual value for the nanoparticle.

The Stefan condition, which is simply an energy balance, is derived in different ways. Obviously one can examine a small change around the front and let the time taken for this change to tend to zero, to provide the differential form. Davis \cite{Davis2001} applies this approach on an arc of surface, taking into account the latent heat release and energy required for creating new surface, hence introducing surface tension. Fedorov and Shulkin \cite{Fedorov2011} simply work with the definition of latent heat in terms of internal energy in the current and bulk configurations to arrive at the effective latent heat  $L_m + \Delta c (T_I-T_m)$. This then replaces the standard latent heat in the Stefan condition. The same expression is obtained by \cite{Alex} through the conservation equations defined in \cite{Bird2007}. If density change between phases is included the conservation route also naturally leads to the inclusion of kinetic energy (with or without curvature) in  \cite{Alex}. Gupta \cite{Gupta} defines a different effective latent heat $L_m + [(\rho_l/\rho_s) c_l - c_s] T_m$ but neglects kinetic energy.
A model including curvature and hence surface energy is derived in \cite{Myers16}, they also go on to show that the term $\Delta c (T_I-T_m)$ plays a very minor role, a result confirmed in \cite{Ribera2016}.

A common issue with studies of phase change is the applied boundary conditions.
The simplest boundary condition to impose in thermal and Stefan problems is the fixed-temperature condition, where the material, which is initially at a constant temperature, has the boundary temperature instantaneously changed to a different, constant value. Clearly this is unphysical and chosen primarily for mathematical simplicity. The work described in  \cite{Font15} showed that, when including density variation, the initial infinite melt velocity  which results from this condition had a significant effect on melt times, and that this effect was more noticeable with smaller particles.  However, even up to the macroscale a discrepancy of approximately 15\% was found between the single- and two-density models for gold particles. In  \cite{Ribera2016} it is shown that when a Newton cooling boundary condition  is employed this discrepancy disappears, even for very high values of the heat transfer coefficient, $h$. Note, they also point out that there is a limit on $h$, beyond which the material would simply vaporise: the mathematical limit $h \rightarrow \infty$, which reduces Newton cooling to the fixed-temperature condition, is not physically possible. Their calculations use the maximum value for $h$. Since it is orders of magnitude larger than typical values the results are perhaps rather forgiving in that they are relatively close to the fixed-temperature results.  In reality much larger differences from the fixed-temperature boundary condition should be observed.

In the following we will discuss phase change with certain variable properties and Newton's law of cooling at the boundary. The Newton condition is more realistic than a fixed-temperature condition, which may be reproduced by introducing an infinite heat transfer coefficient. When a specific configuration is required we will focus on the melting of a spherically symmetric nanoparticle.
The melting of a spherically symmetric nanoparticle is the ideal basic physical situation for developing models of phase change with size-dependent parameters: the symmetry reduces the problem to one dimension, while the geometry requires the introduction of curvature-driven effects. This means that the mathematical model can remain relatively simple while still elucidating the key physical features.

In the following sections we will closely examine the derivation of the Stefan problem via conservation equations and show that they lead to the omission of crucial terms, such as kinetic energy and momentum, in the Stefan condition. These terms have also been missed by researchers following other routes. The correct formulation will then be given.
A main motivation behind this work is the large variety of models and specifically the Stefan condition being applied in recent models. We will demonstrate that what should be a seemingly straightforward derivation can in fact be problematic resulting in missing terms in many of the previously studied models. Even though we will often talk in the context of nanoscale melting the model will be  based on the continuum assumption. The validity of this assumption in the context of phase change has been discussed in detail in \cite{Font13a,Myers2014}, with the basic conclusion that it holds down to around between 2 and 5 nm, depending on the material. The one-phase Stefan problem is a popular simplification, which also has its own significance in the context of ablation, where solid is ablated and subsequently carried away by an external flow \cite{Mitchell08}. With size-dependent parameters the standard form has been shown to lose energy \cite{Evan00,Myers2012,Myers2015}. We will also discuss the one-phase reduction for the current formulation.  Finally, we briefly mention other forms of heat equation specifically designed for very small or very short time scales, which may subsequently be coupled to the new form of Stefan condition.

\section{Governing equations for phase change}\lb{GovSec}

Phase change involves the energy-driven transformation of mass from one form into another.  It is therefore natural to formulate the governing equations in terms of conservation laws and the thermodynamics of a two-phase mixture.
The appropriate conservation equations are provided in \cite{Bird2007}. In the specific context of phase change they are applied in \cite{Alex,Myers16}.

\subsection{Formulation of the bulk equations}
\label{sec:bulk}

The standard governing equations for phase change may be derived from the following conservation laws:
\subeq{
\bea
\lb{conmass}
\pad{\rho}{t} + \nabla \cdot \left(\rho \mathbf{v} \right) = 0 ~ ,\\
\lb{conmom}
\pad{}{t}\left( \rho \mathbf{v}\right)  + \nabla \cdot \left( \rho \mathbf{v}  \otimes \mathbf{v} + p \mathbf{\underline{I}} \right)  = 0 \, ,\\
\lb{conme}
\pad{}{t}\left(\frac{1}{2} \rho v^2\right)  + \nabla \cdot \left(\frac{1}{2} \rho v^2 \mathbf{v} + p \mathbf{v} \right) -p \nabla \cdot \mathbf{v} = 0 \, ,\\
\lb{conen}
\pad{}{t}\left( \rho \left[u + \frac{1}{2} v^2\right] \right)  + \nabla \cdot \left(  \rho \left[u + \frac{1}{2} v^2\right] \mathbf{v} + \mathbf{q} + p \mathbf{v} \right)  = 0 \, .
\eea
}
These represent conservation of mass, momentum, mechanical energy  and total energy, see \cite[\S3.1, 3.2, 3.3, 10.1]{Bird2007}, \cite[\S2.3E]{Alex}. The various parameters are density $\rho$, velocity $\mathbf{v}$ (where $v = |\mathbf{v}|$), pressure $p$, internal energy per unit mass $u$, and the conductive heat flux $\mathbf{q}$.
Equation \eqref{conmass} is a standard mass balance. Equation \eqref{conmom} states that the change in momentum is balanced by the convection of momentum and the effect of pressure.
Note $\otimes$ is the tensor product and $\mathbf{\underline{I}}$ the unit tensor. Since we will be working in three-dimensional space we may define $\mathbf{v} \otimes \mathbf{v}= \mathbf{v} \mathbf{v}^T$, which is a symmetric 3$\times$3 matrix.
Equation \eqref{conmom} therefore differs from the rest in that it is a vector conservation law rather than a scalar form. We will deal with this later.
Equation \eqref{conme} states that the change in kinetic energy is balanced by the input of kinetic energy through the bulk flow, the work done by pressure, $p$, and the rate of reversible conversion to internal energy. Equation \eqref{conen} states that the rate of change of total energy depends on energy flow through convection, conduction and the work done by pressure.

In writing down these equations we have assumed that gravity and viscous effects are negligible. To simplify the discussion in the following we will work in the context of a solid to liquid transition but the model will, of course, be valid for other transitions. We assume that the solid and liquid phases are incompressible, having constant densities $\rho_s$ and $\rho_l$ that, in general, will not be equal. For the equations to hold throughout the material requires an assumption that there is a two-phase mixture or at least a small region, the interface, where there is a transition from solid to liquid. In applying the Rankine--Hugoniot condition below we impose a sharp-interface limit, i.e. the width of the interface tends to zero.

In regions of constant density mass conservation reduces to
\bea
\lb{Divv}
\nabla \cdot   \mathbf{v} = 0 \, .
\eea
Consequently the final term of equation \eqref{conme} is zero.
Subtracting equation (\ref{conme}) from equation (\ref{conen}) leads to the  equation for conservation of internal energy
\bea
\lb{ebal1}
\rho \left( \pad{u}{t}   +  \mathbf{v} \cdot \nabla    u \right) = -  \nabla \cdot  \mathbf{q}     \, .
\eea
The standard heat equation is often derived from this balance by expressing $u$ and $\mathbf{q}$ in terms of the temperature. These are discussed later.

\subsection{Jump conditions at the interface}
\label{sec:jump}

The density and internal energy are discontinuous across a sharp solid-liquid interface.  Conservation of mass and energy links the jump in these two quantities to the motion of the interface. For a scalar conservation law written in divergence form,
\bea
\pad{F}{t} + \nabla \cdot    \mathbf{G}   = 0,
\eea
the Rankine--Hugoniot condition states that the jump in $F$ across a discontinuity moving with normal velocity $v_n$ is
given by
\bea
\label{eqn:RH}
[F]^+_- v_n = [  \mathbf{G} \cdot \hat{\mathbf{n}}]^+_-,
\eea
where $\hat{\mathbf{n}}$ is the unit vector normal to the discontinuity, $v_n$ represents the normal velocity of the interface, and $[\cdot]_{-}^{+}$ denotes the jump. In the following examples the normal vector points into the liquid so that the jump is measured from the solid into the liquid, i.e.,
$[\cdot]_{-}^{+} = [\cdot]_{s}^{l}$. In one-dimensional Cartesian problems the interface is generally denoted by $s(t)$, in one-dimensional cylindrical and radial problems by $R(t)$. In the following we will use either depending on the context.

Equation \eqref{eqn:RH} may be applied immediately to the mass and energy balances to give
\subeq{
\begin{align}
\label{RHm}
(\rho_l - \rho_s) v_n = \left( \rho_l \mathbf{v}_l  - \rho_s \mathbf{v}_s \right) \cdot \hat{\mathbf{n}} &, \\
\rho_l \left( u_l + \frac{v_l^2}{2}\right)(v_n - \mathbf{v}_l\cdot\hat{\mathbf{n}})-\rho_s \left( u_s + \frac{v_s^2}{2}\right)(v_n - \mathbf{v}_s\cdot\hat{\mathbf{n}})
=
\left[(\mathbf{q}_l - \mathbf{q}_s) + (p_l  \mathbf{v}_l - p_s \mathbf{v}_s) \right]   \cdot\hat{\mathbf{n}} &,
\label{RHen}
\end{align}
}
where all variables are evaluated at the solid-liquid interface.   Equation \eqref{RHen} is an extension of the classical Stefan condition accounting for kinetic energy and pressure-volume work. This form of Stefan condition is derived through conservation laws in \cite[\S 2.3E]{Alex}, \cite{Myers2015}.

The momentum balance is a vector relation, so we may not immediately apply the scalar form of Rankine--Hugoniot. A correct form may be obtained by dotting with the unit vector.   If we define   $\mathbf{\underline{R}} = \mathbf{v} \otimes \mathbf{v}$  then, provided $\hat{\mathbf{n}}$ is constant and $\mathbf{\underline{R}}$ symmetric, $(\nabla \cdot \mathbf{\underline{R}}) \cdot \hat{\mathbf{n}}  = \nabla  \cdot ( \mathbf{\underline{R}}\cdot \hat{\mathbf{n}} )$. The appropriate scalar conservation form is then
\bea
\pad{}{t}\left( \rho \mathbf{v} \cdot \hat{\mathbf{n}}\right)  + \nabla \cdot \left[\left( \rho \mathbf{v}  \otimes \mathbf{v} + p \mathbf{\underline{I}} \right) \cdot \hat{\mathbf{n}}  \right] = 0 \,
\eea
and Rankine--Hugoniot may now be applied
\bea
\nonumber
\left[ \left( \rho_l \mathbf{v}_l -   \rho_s \mathbf{v}_s \right) \cdot \hat{\mathbf{n}}\right] v_n &=&  \left[\left( \rho_l \mathbf{v}_l  \otimes \mathbf{v}_l + p_l \mathbf{\underline{I}} \right) \cdot \hat{\mathbf{n}} -
\left( \rho_s \mathbf{v}_s  \otimes \mathbf{v}_s + p_s \mathbf{\underline{I}} \right) \cdot \hat{\mathbf{n}}
 \right] \cdot \hat{\mathbf{n}} \\
 &=&   \rho_l (\mathbf{v}_l \cdot \hat{\mathbf{n}})^2 - \rho_s (\mathbf{v}_s  \cdot \hat{\mathbf{n}})^2 + (p_l -p_s)
 \label{jump_mom}
   \, .
\eea
The second relation is obtained after applying $\mathbf{\underline{I}} \cdot \hat{\mathbf{n}} = \hat{\mathbf{n}}$ and
$[(\mathbf{v}  \otimes \mathbf{v})  \cdot \hat{\mathbf{n}}]\cdot \hat{\mathbf{n}} = (\mathbf{v} \cdot \hat{\mathbf{n}})^2$.

In \cite{Alex} a number of cases of solidification and melting of a semi-infinite one-dimensional bar are described. Following the convention for one-dimensional Cartesian analyses, the interface velocity is written  $v_n = s_t$ and $\mathbf{v}_l \cdot \hat{\mathbf{n}} = v_l$ is the liquid velocity in the $x$-direction. Here we choose an example where the solid is fixed such that $\mathbf{v}_s = \mathbf{0}$. The one-dimensional solid and liquid conductive fluxes are given by $q_s = \mathbf{q}_s\cdot\hat{\mathbf{n}}$ and $q_l = \mathbf{q}_l\cdot\hat{\mathbf{n}}$, respectively. The jump conditions \eqref{RHm}--\eqref{RHen} and \eqref{jump_mom} for mass, momentum and energy then reduce to
\subeq{
\begin{align}
(\rho_l - \rho_s) s_t &=   \rho_l v_l  ,\label{jumprho} \\
\label{jumpmom}
p_s - p_l = \rho_l v_l (v_l-s_t) &= - \rho_s v_l s_t  , \\
\rho_l \left( u_l + \frac{v_l^2}{2}\right)(s_t - v_l )-\rho_s  u_s  s_t
&=
q_l - q_s + p_l  v_l,\label{jumpen}
\end{align}
}
and they hold at $x = s(t)$. Hence a form of the jump conditions given in  \cite[Ch. 2, eqs 25-27]{Alex} is easily retrieved.

The momentum condition \eqref{jumpmom} highlights a problem with this derivation form. It demonstrates that if the density varies between the phases then the
resultant change in velocity causes a change in the pressure. This is fine when dealing with a flat interface.
In the cases of nanowire and nanosphere melting examined in  \cite{Back2014a,McCue2009,Myers2015} the melting proceeds from the outside; following standard notation the interface velocity is denoted $v_n = R_t$ and $\mathbf{v}_l\cdot \hat{\mathbf{n}} = v_l$ is the (purely) radial velocity of the liquid, $\mathbf{v}_s\cdot \hat{\mathbf{n}} =0$ (since the central solid region is fixed). Consequently, the jump conditions which arise for both of these cases will be identical to equations \eqref{jumprho}--\eqref{jumpen} (with $s_t$ replaced by $R_t$).
However, for curved interfaces it is well known that there is also a pressure jump caused by the surface tension
$p_s -p_l = \sigma_{sl} \kappa$, where $\kappa \propto 1/R$ is the curvature and $\sigma_{sl}$ the surface tension between the solid and liquid phases: equation \eqref{jumpmom} misses this contribution.
The problem can be traced to the requirement that the conservation equations hold throughout the whole domain and so in fact represent a single material with varying physical properties, rather than two separate materials. The phase-change interface is the region where properties change and, for a sharp interface, this requires letting its thickness tend to zero, but  this is the limit of a model which contains no interface and so there is no mechanism to account for surface tension.

To correctly account for surface tension in this model a Korteweg stress tensor of the form
\bea
 \mathbf{\underline{K}}  \sim \epsilon (\nabla \phi \otimes \nabla \phi)
\eea
could be introduced into the conservation equations,
where $\epsilon$ is related to surface tension and $\phi$ is the volume fraction of one of the phases. However, this then requires that a form be specified for $\phi$ which correctly reduces to the expected surface tension effect. A simpler `fix' is to add a source term into the conservation equations which accounts for surface tension at the interface:
\subeq{
\bea
\lb{conmom2}
\pad{}{t}\left( \rho \mathbf{v}\right)  + \nabla \cdot \left( \rho \mathbf{v}  \otimes \mathbf{v} + p \mathbf{\underline{I}}\right) + \sigma \kappa\delta(r-s) \hat{\mathbf{n}} = 0 \, ,\\
\lb{conme2}
\pad{}{t}\left(\frac{1}{2} \rho v^2\right)  + \nabla \cdot \left(\frac{1}{2} \rho v^2 \mathbf{v} + p \mathbf{v} \right) -p \nabla \cdot \mathbf{v} + \sigma \kappa \delta(r-s) v_n = 0 \, ,\\
\lb{conen2}
\pad{}{t}\left( \rho \left[u + \frac{1}{2} v^2\right] \right)  + \nabla \cdot \left(  \rho \left[u + \frac{1}{2} v^2\right] \mathbf{v} + \mathbf{q} + p \mathbf{v} \right) + \sigma \kappa \delta(r-s) v_n = 0 \, .
\eea
}
 If an outer surface exists then a second source term may be required (but this will not impact on the derivation of the Stefan condition below).

The Rankine--Hugoniot equations may also be derived by simply integrating the governing equations across a finite region which includes the jump and then letting the thickness of this region tend to zero. Noting that $\int_{-a}^a \delta(x) f(x)\, \ud x = f(0)$ for any $a>0$ and that the velocity $v_n$ may be written $s_t$ in this configuration, the momentum and total energy equations now become
\subeq{
\begin{align}
\lb{pjump} p_s - p_l  &=  \sigma_{sl}\kappa - \rho_s v_l s_t  , \\
\rho_l \left( u_l + \frac{v_l^2}{2}\right)(s_t - v_l )-\rho_s  u_s  s_t
&=
q_l - q_s + p_l  v_l +  \sigma_{sl}\kappa s_t\, ,
\end{align}
}
where all quantities are defined at $r=s(t)$.
The pressure jump now correctly accounts for surface tension, which also affects the energy balance. The addition of surface energy has no effect on condition \eqref{RHm}.
The final equation may be rearranged after invoking equation \eqref{jumprho} to give a form of Stefan condition for a stationary solid phase
\begin{align}
\label{FinStef}
\left[\rho_s \left( u_l-u_s + \frac{v_l^2}{2} \right) - \sigma_{sl}\kappa - \left(1-\frac{\rho_s}{\rho_l}\right) p_l \right]s_t
  &=
    q_l - q_s \, ,
\end{align}
again all quantities are defined at $r=s(t)$. Obviously this may be easily modified for different liquid and solid velocity combinations.

\subsection{Internal Energy and Heat Flux}\lb{IESec}

The Stefan problem is most commonly written in terms of temperature rather than internal energy and heat flux.
The internal energy per unit mass (or specific internal energy) depends on the heat energy and work done against an external pressure.
The heat energy is related to temperature by the heat capacity, $c$, which is defined as the ratio of heat energy entering a system to the temperature change induced by that energy  $c = q/\Delta T$ (if $c$ varies with temperature then we define $q = \int c \, \ud T$). It is standard to discuss heat capacities measured either at constant pressure, $c_{p}$, or constant volume, $c_{v}$. For solids and liquids, which are virtually incompressible, $c_{p} \simeq c_{v}$, so in the following we make no distinction between the two and simply write $c$.  To convert from internal energy to temperature we follow \cite{Alex,Bird2007} and write
\subeq{
\bea
\lb{us}
u_s   &=& c_s (T_s - T_m) - \frac{p_s}{\rho_s} \, , \\
\lb{ul}
u_l   &=& c_l (T_l - T_m) + L_m - \frac{p_l}{\rho_l}\, .
\eea
}
The bulk phase change temperature $T_m$ is measured at a specified reference pressure, denoted $p_\text{ref}$. It is an arbitrary but standard choice to use the bulk value. This choice means that in most Stefan problems, when the temperature reaches the bulk phase change value (for the specified pressure) the enthalpy, $h_s = c_s(T_s-T_m)$, of the solid is zero and the jump between the liquid and solid enthalpies is the latent heat. In
\cite{Alex} a reference internal energy is added to the problem $u_\text{ref} = p_\text{ref}/\rho_s$. However, given that it is a reference state it cancels throughout the calculations hence, for simplicity we will omit this term.

In \cite{Bird2007} a differential form for the internal energy is provided, $\ud u=c_v \ud T$. For incompressible materials the volume is constant in each phase and we may employ this relation in deriving the heat equation.
Note, in \cite{Alex} a slightly different route is followed to obtain $\ud u = c_p \ud T$ (recall subscripts $v$, $p$ denote fixed volume and fixed pressure) despite the fact that pressure may vary in the phases. However, since the phases are incompressible
and so the specific heats are equivalent the result is equivalent and we may write
\bea
\lb{duc}
\ud u = c\, \ud T \, .
\eea

The standard form of heat flux is based on Fourier's law
\bea
\lb{Fq}
\mathbf{q} = -k \nabla T \,
\eea
where $k$ is the thermal conductivity. However, there are situations where this does not hold.
These are described in more detail in \S \ref{NonFSec}.

The heat equations may be obtained from \eqref{ebal1} by noting that $\ud u = c\, \ud T$   and using the appropriate law for the heat flux, in this case \eqref{Fq}. Since the material properties  are  constant within each phase  we obtain
\begin{align}
\rho_s  c_s \pad{T_s}{t} &=  k_s \nabla^2 T_s   ,  &0 < \;\; &x < s(t)\,  \,, \label{SolEn} \\
\rho_l c_l \left( \pad{T_l}{t} + \mathbf{v_l} \cdot \nabla T_l\right) &=  k_l \nabla^2 T_l,  &s(t) < \;\; &x  \, \, ; \label{LiqEn}
\end{align}
$s(t)$ is replaced by $R(t)$ in cylindrical and spherical problems.
These equations may also be obtained by substituting for $u$, via equations (\ref{us}, \ref{ul}), into equation \eqref{ebal1}. This leads to terms of the form $\ud (p/\rho)/\ud t$. In  \cite[ch. 11.2]{Bird2007} these are shown to be zero for incompressible phases.
Using definitions (\ref{us}, \ref{ul}) the Stefan condition \eqref{FinStef} may now be written, with the help of \eqref{pjump}, as
\begin{align}
\label{FinStef2}
\rho_s\left[   (c_l-c_s) (T_I-T_m)+ L_m    - \frac{1}{2}\left(1-\left(\frac{\rho_s}{\rho_l}\right)^2\right)  s_t^2 \right]s_t
&= k_s \nabla T_s\cdot\hat{\mathbf{n}} - k_l \nabla T_l\cdot\hat{\mathbf{n}}  \, ,
\end{align}
where $T_I$ is the interface temperature (not necessarily equal to $T_m$). Equations \eqref{SolEn}--\eqref{FinStef2} define the two phase Stefan problem when there is a density difference between the phases and a stationary solid.

In many studies
kinetic energy is neglected in the Stefan condition, either because it is small compared to the latent heat or because density variation is not considered. However, it is worth pointing out that when applying a fixed-temperature boundary condition, $T_s(0,t)=T_0$, where $T_0$ is constant, then the initial solidification velocity is infinite and kinetic energy dominates. Also in the final stages of melting a cylinder or spherical material the melt rate tends to infinity. Hence there are times when kinetic energy is dominant but it is omitted  from virtually all analyses, see \cite{Font15}.
Another  important point to note is that equation \eqref{FinStef2} does not match the form provided in the textbook \cite{Alex}. We discuss this now since their equations have been taken up by subsequent authors, see \cite{Font15,Ribera2016,Schulte18} for example. The issue may be traced to the solution of their Problem 24, \S 2.3. Their equation (48a) defines the internal energy in the solid at the interface as the reference value, this requires choosing $p_s(s) =  p_\text{ref}$. To derive equation (48b) they neglect a term in the interface value of the internal energy of the liquid, to do this requires $p_l(s) =  p_\text{ref}$, i.e. $p_l(s) = p_s(s)$. Yet their equation (33) defines the pressure jump $p_l(s)-p_s(s) \ne 0$. The goal of their equation (37) is to determine $p_s(s) - p_\text{ref}$ which then introduces an erroneous kinetic energy term, so affecting the kinetic energy in the final equation and thus all expressions of subsequent authors who take this as their base model.

\section{Application to standard configurations}
\subsection{One-dimensional solidification of a semi-infinite bar}

To illustrate the method we now focus on the standard problem of the solidification of a semi-infinite one-dimensional bar driven by cooling at $x=0$.
The solid region occupies $x \in[0,s(t))$, the liquid occupies $x > s(t)$, where $s(0)=0$. In this scenario it makes sense to keep the solid material fixed and allow the fluid to move due to the density change, so that $\mathbf{v}_s=\mathbf{0}$ and then we may apply equations \eqref{SolEn}--\eqref{FinStef2} without any amendment for a moving solid. The interface velocity is written $v_n = s_t$.

The liquid energy equation \eqref{LiqEn} requires knowledge of the velocity $v_l(x,t) = \mathbf{v}_l \cdot \hat{\mathbf{n}}$, where $\hat{\mathbf{n}} = \hat{\mathbf{x}}$. In general this may be obtained via the mass conservation equation \eqref{Divv}, which here reduces to $\partial v_l / \partial x = 0$ and hence $v_l(x,t) = v_l(t)$. Since the liquid velocity is independent of $x$ we only require its value at a single position to determine the value throughout the liquid.  It may be calculated from the jump condition at $x=s(t)$, equation \eqref{RHm},
\bea
\lb{vlCart}
v_l(t)  = \left( 1-\frac{\rho_s}{\rho_l}\right) \nd s t \, .
\eea
This in turn depends on the interface velocity, $s_t$, which is determined by the Stefan condition \eqref{FinStef2}.
The problem formulation is completed by specifying appropriate initial and boundary conditions.
Here an initial temperature must be specified in the liquid, the solid does not exist at $t=0$ so requires no initial condition, while the phase change front satisfies, $s(0)=0$. In the case of melting we would specify an initial solid temperature and no liquid region.
At $x=0$ any of the standard conditions may be applied. In the far-field the liquid  temperature approaches the initial value. At the moving boundary the temperature depends on the situation. Most commonly the interface temperature is set to the bulk melt temperature $T_I=T_m$, so simplifying the Stefan condition. However, even with a flat interface $T_I$ may vary. For example when dealing with the solidification of supercooled materials \cite[\S 2.4F]{Alex}, \cite{Myers2012} there exists a nonlinear relation between the degree of supercooling and the front velocity \cite{Ashby06,Font13b}
\bea
\nd s t = C_0 \ue^{-C_1/T_I} (T_m - T_I) \, ,
\eea
where $C_i$ are constants. This equation determines the phase change temperature.
A linearised version, valid for small supercooling, is more standard in mathematical studies \cite{Alex,Font13b}
\bea
T_I = T_m - \Phi s_t ~ .
\eea
The variation of $T_I$ with curvature is discussed in the following section.

With the above definitions the solidification of a semi-infinite one-dimensional bar may now be carried out by appropriate numerical or approximate methods.

\subsection{Spherically symmetric nanoparticle melting}
\label{sec:sphere}

\begin{figure}
\centering
\includegraphics[width=0.4\textwidth]{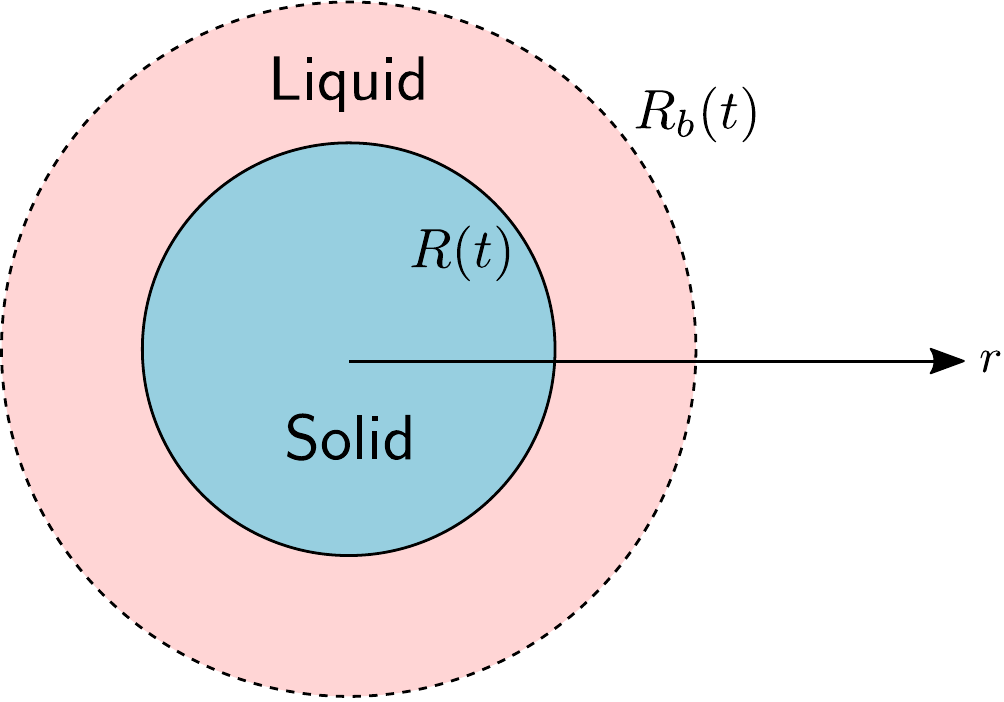}
\caption{Spherically symmetric nanoparticle melting.  The regions of solid and liquid phase are defined by $0 < r < R(t)$ and $R(t) < r < R_b(t)$, respectively, where $r$ is the radial coordinate and $t$ is time.  The melting process is driven by an influx of thermal energy from the surrounding environment into the liquid phase.}
\label{fig:np}
\end{figure}

A popular problem in the literature is  that of  spherically symmetric nanoparticle melting. We now examine this since it  permits a discussion of the model reduction in spherical co-ordinates and also introduces size-dependent parameters. The physical situation is depicted in Figure \ref{fig:np}.

The nanoparticle is envisioned as being surrounded by a hot ambient environment with temperature $T_a$.  As the solid  melts, two interfaces evolve; these are the solid-liquid and the liquid-air.  Due to the density difference between the solid and liquid, the liquid-air interface will grow in order to accommodate the change in volume that occurs during melting.   The solid and liquid  occupy the  domains given by $0 < r < R(t)$ and $R(t) < r < R_b(t)$.  In this geometry, the velocity of the solid-liquid interface is denoted $v_n = \ud R / \ud t$ and the unit normal vector to the interface is $\hat{\mathbf{n}} =\hat{\mathbf{r}}$.

The heat equations are defined by the spherically symmetric versions of \eqref{SolEn}--\eqref{LiqEn}. As before an expression for the liquid velocity is obtained first by integrating the mass conservation equation \eqref{Divv} which indicates  $v_l(r,t) = A(t)/r^2$. The function $A(t)$
is determined by the jump condition \eqref{RHm} at $r=R(t)$
\bea
\lb{vlR}
(\rho_l - \rho_s) \nd R t = \rho_l v_l(R,t) ~ .
\eea
Hence we find the relation between the liquid velocity and the rate at which the front moves
\begin{align}
\lb{vleq}
v_l(r,t) = \left(1-\frac{\rho_s}{\rho_l}\right) \frac{R^2}{r^2} \nd R t \, .
\end{align}
This specifies the liquid  velocity in equation \eqref{LiqEn}.

At the outer boundary the liquid velocity is simply $v_l(R_b,t) = \ud R_b/\ud t$.
Noting that if we set $r=R_b$, $v_l(R_b,t) = \ud R_b/\ud t$ in \eqref{vleq} and integrate then the following relation is obtained
\begin{align}
\lb{Rbeq}
R_{b}^3= \frac{\rho_s}{\rho_l} R_0^3+ \left(1- \frac{\rho_s}{\rho_l}\right) R^3 ~ ,
\end{align}
where we have imposed $R_b(0)=R(0)=R_0$. This determines the position of the outer boundary.

The interface velocity is defined by the Stefan condition \eqref{FinStef2}.

The model of nanoparticle melting is completed by imposing initial and boundary conditions.
Boundedness of the temperature at the origin requires
\begin{align}
\pad{T_s}{r}\bigg|_{r = 0} = 0.
\end{align}
At the solid-liquid interface, the solid and liquid temperatures are equal to the phase change temperature,
\begin{align}
  T_s(R,t) = T_l(R,t) = T_I(R) ~ .
\end{align}
At a curved interface (in the absence of supercooling) the temperature may be approximated by the Gibbs-Thomson relation
\bea
T_I =   T_m\left(1 - \frac{\sigma_{sl}\kappa}{\rho_s L_m}\right) \,
  \label{eqn:GT}
\eea
with curvature $\kappa$. This may be viewed as a limit of the generalised Gibbs-Thomson which accounts for energy contributions due to changes in pressure and specific heat between the phases, see \cite{Alex,Font13a}. If $\kappa$ is sufficiently small then the relation reduces to $T_I=T_m$.

Since this is a finite domain a thermal condition is also required at the outer boundary.
Thermal energy is gained from the surrounding material, proportional to the difference in temperatures across the boundary,
\bea
\nd {E_n} {t} = 4\pi R_b^2 h [T_a -T_l]_{r=R_b} \, ,
\eea
where $T_a$ is the temperature of the surrounding material and the constant of proportionality $h$ is the heat transfer coefficient. However energy is used up by the resistance to the change in particle size carried out on an environment which has constant pressure $p_a$
\bea
\nd {E_s} {t} = -4\pi R_b^2 p_a \left. v_l\right|_{r=R_b}\, .
\eea
Just inside the particle the change in energy is specified by
\bea
\lb{ETins}
\nd E t = 4\pi R_b^2 [-q_l - p_l v_l + \sigma_{sl} v_l]_{r=R_b} \,
\eea
(equation \eqref{ETins} is derived in the following section).
Conservation of energy requires these three expressions to balance, leading to
\bea
- (q_l +p_l v_l- \sigma_{la} \kappa v_l) = h (T_a -T_l) -p_a v_l
\eea
at $r=R_b(t)$. The Young-Laplace equation states that the pressure jump across the interface is given by $p_l -p_a = \sigma_{la} \kappa$, this removes all terms involving $v_l$. Then, despite the pressure work and creation of new surface the standard Newton cooling condition is retrieved
\begin{align}
q_l = - h (T_a - T_l) \, , \quad r = R_b(t) \, .
\label{bc:newton}
\end{align}
Letting $h \ra \infty$ leads to the standard fixed-temperature boundary condition $T_l(R_b) = T_a$.

The initial radius of the nanoparticle is denoted as $R(0) = R_b(0) = R_0$.   An initial solid temperature must be specified, in the literature this is often taken as the melt temperature $T_s(r,0) = T_I(R_0)$. In this way melting starts at $t=0$ and there is no need to investigate an initial heating phase. Due to the lack of a liquid phase at $t = 0$, an initial condition for the liquid temperature is not required. However, a small-time analysis can be used to calculate the liquid temperature for arbitrarily small times, which may then be applied in a numerical solution; see Appendix \ref{sec:small_time}.

\section{Energy conservation}

\label{sec:energy}
The derivation of the previous governing equations was based on conservation equations and consequently energy is naturally conserved, provided exchange at an outer boundary is correctly accounted for. However, a common simplification of the Stefan problem is the one-phase reduction. It has been demonstrated that in the presence of melting point depression the standard reduction does not conserve energy. We will now detail the energy conservation argument to show that energy loss or gain only occurs at the boundary. The argument will then inform the discussion of the one-phase reduction of \S \ref{sec:one_phase} and will also clarify the expression for the boundary condition \eqref{ETins}.

To verify that energy is conserved, we note that the governing equations have been derived from conservation laws, even the Stefan condition denotes conservation across the interface, so energy conservation should be both automatic and obvious. We may see this explicitly by considering the total energy in the spherically symmetric system. If we use the standard form of total energy conservation, equation \eqref{conen} (which does not account for interfaces), then the surface energy must be added at both the exterior and melt front. The total energy may then be written
\bea
\lb{EBal}
\frac{E}{4\pi} =  \int_0^R \rho_s u_s r^2 \ud r +  \int_R^{R_b} \rho_l\left( u_l + \frac{v_l^2}{2}\right) r^2 \ud r + R^2 \sigma_{sl} + R_b^2 \sigma_{la}\, ,
\eea
where the terms in the integrals obviously coincide with those in the total energy expression of equation \eqref{conen} (after taking into account the fact that the solid velocity is zero).
This equation may also be derived, without explicitly adding the surface energy terms via equation \eqref{conen2} after including a second source term to account for the outer boundary.

The rate of change of energy must balance that passing through the boundary. Taking the derivative of \eqref{EBal} with respect to time gives
\bea
\begin{split}
\frac{1}{4\pi}\nd{E}{t} =& \int_0^R \rho_s \pad{u_s}{t} r^2 \ud r + \rho_s u_s|_{r=R} \,R^2 \nd{R}{t}  + \int_R^{R_b} \rho_l\pad{}{t}\left( u_l + \frac{v^2}{2}\right) r^2 \ud r  \\ & +
\rho_l\left[ u_l + \frac{v^2}{2}\right]_{r=R_b} R_b^2 \nd{R_b}{t}-\rho_l\left[u_l + \frac{v^2}{2}\right]_{r=R} R^2\nd{R}{t} +2 \sigma_{sl} R \nd{R}{t} +2 \sigma_{la} R_b \nd{R_b}{t}\, .
\end{split}
\eea
We may use equation \eqref{conen}, with $\mathbf{v}_s = \mathbf{0}, \mathbf{v}_l = v_l \hat{\mathbf{r}}$, to remove the time derivatives inside the integrals
\bea
\begin{split}
\frac{1}{4\pi}\nd{E}{t} =& - \int_0^R  \pad{ }{r} (r^2 q_s) dr - \int_R^{R_b} \pad{}{r}\left[r^2 \left( \rho_l v_l \left(u_l + \frac{v_l^2}{2}\right) +q_l + p_l v_l \right) \right] \ud r  + \rho_s u_s|_{r=R}  \,R^2 \nd{R}{t} \\  & +
\rho_l\left[ u_l + \frac{v_l^2}{2}\right]_{r=R_b} R_b^2 \nd{R_b}{t}
 -\rho_l\left[ u_l + \frac{v_l^2}{2}\right]_{r=R} R^2\nd{R}{t} +2 \sigma_{sl} R R_t +2 \sigma_{la} R_b \nd{R_b}{t} \, .
\end{split}
\eea
Now we express $v_l(R,t)$ in terms of $\ud R/\ud t$, via \eqref{vlR}, and $v_l(R_b,t)=\ud R_b/\ud t$. In spherical co-ordinates we may write $\kappa r^2 = 2 r$ then after some algebra the above expression may be reduced to
\bea
\frac{1}{4\pi}\nd{E}{t} &=& R^2 \left[q_l -q_s  - \left\{\rho_s \left( u_l -u_s + \frac{v_l^2}{2} \right) - \sigma_{sl}\kappa- \left(1-\frac{\rho_s}{\rho_l}\right)p_l \right\} \nd{R}{t} \right]_{r=R} \nonumber \\ &-& R_b^2 \left[q_l + (p_l -\sigma_{la}\kappa)\nd{R_b}{t} \right]_{r=R_b} ~ .
 \lb{EtRed}
\eea
The terms in the square brackets evaluated at $r=R$ are simply the Stefan condition \eqref{FinStef}
and may therefore be set to zero leaving
\bea
\lb{EBalFin}
\frac{1}{4\pi} \nd{E}{t} = -  R_b^2 \left[q_l + (p_l - \sigma_{la} \kappa) \nd{R_{b}}{t} \right]_{r=R_b} \, .
\eea
So, energy only leaves or enters the system at the outer boundary. The change is a result of  the heat flux there, the work done against the ambient pressure to change the size of the particle and also the  creation of new surface. This balance is the appropriate  form required in the derivation of the Newton cooling boundary condition of the previous section.

\section{Model reductions}

\subsection{Non-dimensionalisation}
\label{sec:nondim}
The relative importance of the terms in the Stefan problem can be assessed by casting the model into dimensionless form.
We define
\bea
t = \tau \hat{t} \qquad x = L \hat{x} \qquad T = T_m + \Delta T \hat{T}
\eea
where  the time, length and temperature scales are $\tau, L, \Delta T$ respectively. These choices are not unique and should  be based on the physical situation. A standard time-scale comes from writing the liquid heat equation in non-dimensional form. Taking the spherical version (and immediately dropping the hat notation) we obtain
\bea
\frac{\rho_l c_l L^2}{k_l \tau} \left(\pad T t + \frac{\tau V}{ L} v_l \pad T r \right) = \frac{1}{r^2}\pad{}{r}\left(r^2 \pad T r \right)
\eea
and consequently set $\tau = \rho_l c_l L^2/k_l = L^2/\alpha_l$, where $\alpha_l$ is the thermal diffusivity of the liquid, and the velocity scale $V=L/\tau$. An obvious length-scale would be the initial particle radius, $L=R_0$, while, since the phase change is driven by the external temperature, we choose $\Delta T = T_a - T_m$. With these scalings we obtain
\bea
\lb{NDheat}
\pad{T_s}{t}   = \alpha  \frac{1}{r^2}\pad{}{r}\left(r^2 \pad {T_s}{ r} \right) ~, \qquad \pad {T_l} {t} +  v_l \pad {T_l}{ r}  = \frac{1}{r^2}\pad{}{r}\left(r^2 \pad {T_l}{ r} \right)
\eea
where $\alpha=\alpha_s/\alpha_l$.
With the flux described by Fourier's law the Stefan condition becomes
\bea
\lb{NDStef}
\rho \beta \left[ 1- \frac{\gamma \Gamma}{R} - \frac{\delta}{2}  R_t^2 \right] \nd R t = k \pad{T_s}{r} - \pad{T_l}{r}
\eea
where $\beta = L_m/(c_l \Delta T)$ is the Stefan number (or inverse Stefan number, both forms are common throughout the literature), $\gamma = (c_l-c_s)\Delta T/L_m$,  $\Gamma = 2 T_m\sigma_{sl}/(\rho_s L_m R_0 \Delta T)$, $\rho=\rho_s/\rho_l$ and $\delta =   R_0^2\left(1-\rho^2\right) /(L_m \tau^2)$. If the Stefan number is large equation \eqref{NDStef} indicates $R_t = \Oc(1/\beta) \ll 1$ and the phase change is slow. To focus on the evolution of the particle a new time variable $t^* = \hat{t}/\beta= t/(\beta \tau)$ may be used.

The dimensionless position of the outer boundary is given by
\begin{align}
R_b = \left(\rho - R^3 (\rho - 1) \right)^{1/3} \label{eqn:nd_Rb}
\end{align}
and the velocity of the liquid is
\begin{align}
v_l(r,t) = (1 - \rho)\frac{R^2}{r^2}\nd{R}{t}. \label{eqn:nd_vl}
\end{align}
The boundedness condition at the origin remains unchanged.
The Gibbs-Thomson equation becomes $T_I = -\Gamma / R(t)$ and
\begin{align}
T_s(R(t),t) = T_l(R(t),t) = T_I(t) ~.
\end{align}
The Newton condition may be written as
\begin{align}
\left.\pad{T_l}{r}\right|_{r = R_b(t)}  = Nu [1 - T_l(R_b(t),t)],
\label{eqn:nd_newton}
\end{align}
where $Nu = h R_0 / k_l$ is the Nusselt number. The initial conditions are
\begin{align}
T_s(r,0) = T_I(0)= -\Gamma, \quad R(0) = R_b(0) = 1.
\label{ic_nd}
\end{align}

The non-dimensional numbers provide information about the physical process as well as identifying dominant and negligible terms. In Table \ref{tab:tin} typical parameter values are presented for tin and gold, these are taken from \cite{Font13a,Font15,Ribera2016}. The value for $h_\text{max}$ for gold is calculated from $h_\text{{max}}= c_s \sqrt{\rho_s B/3}$, where $B=220 \times 10^9$~Pa is the bulk modulus, see \cite{Ribera2016}. Four numbers are independent of the applied temperature difference:
using the values for tin shows
\[
\tau \approx (6.2 \times 10^4 ~ \text{s/m}^{2}) R_0^2 ~, \quad \gamma \Gamma \approx \frac{10^{-10} ~ \text{m}}{R_0} ~ , \quad  \delta \approx \frac{3\times10^{-16} ~ \text{m}^2}{R_0^2} ~ , \quad  Nu \approx (1.6 \times 10^8 ~ \text{m}^{-1}) R_0 \, .
\]
With a 10~nm sphere this indicates
a heat conduction times-scale on the order $10^{-12}$~s. The term $\gamma \Gamma$ represents the effect of the melting point variation on the Stefan condition.  Only in the very final stages of melting will this play a significant role: when $R = \ord{1}$~nm we expect  this to cause a variation of the order 10\%. This finding is in keeping with that reported in \cite{Myers16}. The kinetic energy term $\delta$ is order unity and so may play an important role, in keeping with the findings of \cite{Font15}. This is especially true when combined with a fixed-temperature boundary condition. The Nusselt number $Nu$ is also of order unity, but this is when employing the maximum possible heat transfer coefficient. In general $h \ll 10^9$~W/(m$^2$$\cdot$K) and we would expect the Nusselt number to be negligible, that is, it appears sufficient to apply a zero flux condition $\partial T_l/\partial r = 0$ at the outer boundary. However, this is not the whole story, clearly to start the melting process some input of heat is required, i.e. a non-zero flux. So either the Nusselt number term must be retained or a rescaling applied at least for small times. Once melting begins the fact that the melt temperature of the solid decreases with decreasing radius as does the surface area to be melted  means that the energy required to melt the front also decreases. This suggests a runaway process which, once started, will only increase in speed. This may be seen in all published results for the radius of a melting nanoparticle where $R_t \ra -\infty$ as $R\ra 0$, see \cite{Back2014b,Back2014a,Back2014,Font13a,Font15,McCue2009,Ribera2016,Wu2009} for example. Similarly the analysis of \cite{Hennessy2018}, using the Maxwell-Cattaneo heat equation, shows `supersonic melting', where the speed of the melt front is faster than the speed of heat propagation.

The remaining non-dimensional numbers depend on the temperature scale, when $\Delta T = 10$~K:
\bea
\beta \approx 22 ~ , \quad \Gamma \approx \frac{1.5\times 10^{-8} ~ \text{m}}{R_0} ~. \nonumber
\eea
The melting time-scale is $\beta \tau$, which is of order $10^{-11}$~s, i.e. 10 picoseconds.
The Stefan number $\beta$ may be considered large, indicating that a pseudo-steady solution will provide good accuracy. This is discussed in Section \ref{sec:large_beta}. The difference between the bulk melt temperature and the initial temperature is represented by $\Gamma$ which is here of order unity.

For gold we find similar values and hence the same conclusions: the main difference coming through a larger kinetic energy contribution due to a greater density difference between solid and liquid phases.

\begin{table}
\centering
\caption{Values of the physical parameters for tin and gold.}
\label{tab:tin}
\begin{tabular}{|c|c|c|c|}
\hline
 & Symbol & Tin & Gold \\
\hline
Density of liquid kg/m$^3$ & $\rho_l$ & 6980 & 17300\\
Density of solid kg/m$^3$ & $\rho_s$ & 7180 & 19300\\
Thermal conductivity of liquid (bulk) W/(m$\cdot$K) & $k_l$ & 30 & 106 \\
Thermal conductivity of solid (bulk) W/(m$\cdot$K) & $k_s$ & 67 & 317 \\
Specific heat of liquid J/(kg$\cdot$K) & $c_l$ & 268 & 163 \\
Specific heat of solid J/(kg$\cdot$K) & $c_s$ & 230 & 129\\
Melt temperature (bulk) K  & $T_m$ & 505 & 1337 \\
Latent heat (bulk) J/kg & $L_m$ & 58500 & 63700 \\
Solid-liquid surface energy N/m & $\sigma_{sl}$ & 0.064 & 0.27  \\
Vapour-liquid surface energy N/m  & $\sigma_{la}$ & 0.589 & 1.15 \\
Maximum heat transfer coefficient W/(m$^2\cdot$K) & $h_\text{max}$ & $4.7\times 10^{9}$ & $4.9\times 10^{9}$ \\
\hline
\end{tabular}
\end{table}

\subsection{One-phase models}
\label{sec:one_phase}

A common simplification in the study of Stefan problems is the one-phase reduction, whereby one of the phases is neglected.
In the absence of supercooling or size-dependent properties this reduction may be achieved by simply setting the temperature in the neglected phase to the bulk melt temperature. The heat equation is then automatically satisfied while the corresponding conduction term in the Stefan condition is zero. However, when supercooling or size dependence occurs this reduction does not conserve energy, see \cite{Evan00,Myers2012}.

Taking into account both surface-tension effects and supercooling, an energy-conserving formulation based on negligible conductivity in the solid phase, i.e. $k_s \ll k_l$, is derived in \cite{Evan00}. They highlight a number of past papers where incorrect reductions are made, for example by naively setting $k_s =0$.
Wu \emph{et al} \cite{Wu2009} support the assertions of \cite{Evan00} going on to state that if the initial temperature is different to the phase change temperature then the one-phase limit may only be derived under the assumption $k_s \ll k_l$.
Due to the way heat is conducted through a material the thermal conductivity of a solid is nearly always greater than in its liquid phase: it is hard to conceive of a material where $k_s \ll k_l$.  In \cite{Myers2012} an asymptotic reduction is derived which is valid for the more physically sensible  case $k_s > k_l$. Consequently the energy-conserving one-phase reduction in the limit $k_s \gg k_l$ is provided in
\cite{Myers2012}.

In the context of nanoparticle melting the issue of one-phase reductions is discussed in more detail in \cite{Myers2015} where it is shown that previous problems arise due to an inconsistency in assigning temperatures. Since most authors start from the standard governing equations they are often unaware that continuity of temperature at the interface has already been imposed.  If the one-phase reduction is then made by setting $T_s$ to some value not equal, at all times, to the temperature imposed during the derivation  then the inconsistency manifests itself in a loss of energy. It is concluded that energy-conserving forms may be written down immediately, without resorting to asymptotic reductions, provided consistency is maintained.

A one-phase model can be systematically derived from the non-dimensional system \eqref{NDheat}--\eqref{NDStef} after assuming $\rho c = (\rho_s c_s)/(\rho_l c_l) = \ord{1}$ and $k = k_s/k_l \gg 1$. Expanding in terms of $1/k \ll 1$ the solid temperature in a melting spherical nanoparticle is given to first order by
\bea
\lb{Ts1p}
T_s(r,t) = T_I(R(t)) + \frac{1}{k} \frac{\rho c}{6} \pad{T_I}{t} (r^2-R^2) \, ,
\eea
where boundedness at $r=0$ has been applied.
The leading-order term $T_I$ reflects the fact that the solid is a good conductor and hence rapidly transmits the boundary temperature. The Stefan condition states that the melt front is driven by the heat flux in the solid and liquid. With Fourier's law the heat flux is proportional to the temperature gradient, which is zero to leading order in the solid since $T_I$ is a function of time. However, in equation \eqref{NDStef} the solid temperature gradient is multiplied by $k$ meaning that the first-order term enters the leading-order balance and must be accounted for. The appropriate Stefan condition is then
\bea
\label{1PStef}
\rho \beta \left[ 1+ \gamma T_I - \frac{\delta}{2}  R_t^2 - \frac{c \Gamma}{3R } \right] R_t =  - \pad{T_l}{r} \, .
\eea
The final term in the square brackets derives from the energetic contribution of the solid to the melting process.

Given that the above analysis involves a varying solid temperature the term one-phase reduction may not be strictly appropriate. Taking only the leading order temperature $T_s = T_I(R(t))$ would be more appropriate for a true one-phase reduction, but this removes the $\Gamma$ term (i.e. the fourth term) from the Stefan condition which is at leading order, rather than a small correction. Further, the leading-order reduction for the temperature only satisfies the heat equation in the limit $k \ra \infty$, so the form given by \eqref{Ts1p}--\eqref{1PStef} is preferable.

The main message of \cite{Myers2015} is that energy conservation with the one-phase reduction is largely an issue of consistency, with  problems arising due to the use of accepted forms for the heat equations and Stefan condition without taking into account the assumptions applied during their derivation. With the current derivation
energy conservation is not an issue. The Stefan condition \eqref{FinStef} does not specify a form for the temperature hence the energy balance \eqref{EBalFin} holds, independent of the form of temperature. Depending on the choice of temperature an inconsistency can arise when trying to satisfy the heat equation. The choice of a constant temperature $T_s=T_m$ satisfies the heat equation, but not the boundary condition at $r=R$.
The choice of $T_s$ specified by \eqref{Ts1p} is accurate to $\ord{1/k^2}$ while $T_s=T_I(R(t))$ only satisfies the heat equation in the limit $k \ra \infty$. All consistent choices can conserve energy provided the Stefan condition is used in the form \eqref{FinStef}. A common error in a number of papers is the use of the form \eqref{FinStef2}, which is based on setting the interface temperature $T_s=T_l=T_I(t)$: the subsequent choice $T_s = T_m$ then leads to loss of energy conservation.


\subsection{Limit of large Stefan number}
\label{sec:large_beta}

The Stefan number characterises the
relative time scales of melting and thermal diffusion. The standard large Stefan number expansion, which corresponds to slow melting, requires a re-scaling of (dimensionless) time such that $t = \beta t^*$ and reduces the governing equations  to
\subeq{
\begin{align}
\label{solid-rho1}
\frac{1}{\beta}\pad{T_s}{t}
&= \frac{k}{r^2} \frac{\partial}{\partial r} \left( r^2 \frac{\partial T_s}{\partial r} \right), &0 < \;\; &r < R(t), \\
\label{liquid-rho1}
 \frac{1}{\beta}\left(\pad{T_l}{t} + v_l \pad{T_l}{r}\right)  &= \frac{1}{r^2} \frac{\partial}{\partial r} \left( r^2 \frac{\partial T_l}{\partial r} \right), &R(t) < \;\; &r < R_b(t) ~,
\end{align}
}
where now $v_l = (1-\rho)(R^2/r^2)R_{t}$ and the star notation has been dropped.
The Stefan condition becomes
\begin{align}\label{large_beta_Stef}
\rho \left[1+ \gamma T_I - \frac{\delta}{2\beta^2} R_{t}^2 \right] \frac{\ud R}{\ud t} = k \frac{\partial T_s}{\partial r}\bigg|_{r=R} - \frac{\partial T_l}{\partial r}\bigg|_{r=R} ~.
\end{align}
The coefficient $1/\beta$ in the heat equations indicates that the system is now pseudo-steady, that is the heat equations at leading order assume their steady form indicating that the growth is slow compared to heat conduction. Time enters through the position of the moving boundary, specified by equation \eqref{large_beta_Stef}.

The leading-order temperatures are
\begin{subequations}
\label{large_beta_T}
\begin{align}
	T_s(r,t)&=T_I(t) ~,\\
    T_l(r,t)&= \frac{A(t)}{r} + B(t),
	\end{align}
\end{subequations}
where $A(t)$, $B(t)$ are
\begin{equation}
  A(t)=\frac{Nu R R_b^2 (T_I-1)}{NuR_b (R_b-R)+R} ~, \qquad B(t) = T_I-\frac{A}{R} ~ .
  \label{large_beta_AB}
\end{equation}

The kinetic energy term in the Stefan condition now has a factor $1/\beta^2$ indicating it may be neglected. However, it should be treated with care. In the standard problem where the boundary temperature is fixed $T_l(R,t) = T_a$ the initial front velocity $R_{t}$ is infinite. When the particle size tends to zero, with all standard boundary conditions $R_{t} \ra \infty$. In both cases kinetic energy may be dominant, indicating the need for re-scaling. Even with a Newton boundary condition the reduction may not be trivial, as discussed below.

Provided $R_t \ne \ord{\beta}$
substituting the temperatures \eqref{large_beta_T} into the Stefan condition \eqref{large_beta_Stef} yields a nonlinear differential equation for $R$. At leading order this is
\begin{equation}
\label{ODE_largebeta}
	\nd{R}{t}=\frac{A(t)}{\rho R^2(1+\gamma T_I(t))},
\end{equation}
subject to the initial condition $R(0)=1$. Note, the limit $Nu \ra 0$ indicates $A \ra 0$ and hence $R_{t} \ra 0$ as discussed in the previous section. With a fixed-temperature boundary condition the initial front velocity is infinite, in the examples studied later we use the highest value of heat transfer possible so leading to the possibility of high velocities. Here we note that $A(0) = -Nu(1 + \Gamma)$, which from \eqref{ODE_largebeta} implies that the initial speed of the interface is given by $R_t(0) = -\rho^{-1}Nu(1 + \Gamma)(1-\gamma\Gamma)^{-1}$. The role of interface kinetic energy can then be estimated from the magnitude of
\begin{align}
    K = \frac{\delta}{2\beta^2} [R_t(0)]^2 = -\frac{\delta (Nu)^2(1+\Gamma)^2}{2\beta^2 \rho^2(1 - \gamma \Gamma)^2}.
    \label{eqn:ke_estimate}
\end{align}
The quadratic factors of the Nusselt number $Nu$ and $1 + \Gamma$ that appear in the numerator of \eqref{eqn:ke_estimate} imply that heat transfer and melting point depression, both of which are related to the size of the nanoparticle, can amplify the effect of kinetic energy. In the case of a 10~nm gold nanoparticle subject to a heating of $\Delta T = 10$~K, we find that $\beta \simeq 39$, $\delta \simeq -54$, $Nu \simeq 0.46$, and $\Gamma \simeq 5.9$, leading to $K \simeq 0.15$.  This value of $K$ is not particularly small, especially compared to $\beta^{-1} \simeq 0.03$, the small number used in the asymptotic expansion that led to \eqref{ODE_largebeta}. Consequently we anticipate inaccuracy in the expansion (which we will see in the numerical results). The problem could be dealt with via a more thorough analysis involving re-scaling near the boundary and then matching to an outer expansion. However, the issue really arises since we have chosen such a high value of heat transfer coefficient, in general this high velocity should not be an issue.
Consequently we take a much simpler solution, by leaving kinetic energy in the Stefan condition and writing
  \begin{align}
    \rho\left[1 + \gamma T_I(t) - \frac{1}{2}\frac{\delta}{\beta^2}\left(\nd{R}{t}\right)^2\right]\nd{R}{t} = \frac{A(t)}{R^2}.
    \label{ODE_largebeta_ke}
  \end{align}
The kinetic energy term enters plays a significant role when $K$ is large, as $K$ reduces its effect will become less noticeable.
The solutions for the temperature remain the same and are given by \eqref{large_beta_T}--\eqref{large_beta_AB}.
Numerically solving  equation \eqref{ODE_largebeta_ke} is relatively straightforward but requires dealing with the cubic equation for $\ud R/\ud t$ at each time step (rather than the simpler linear equation when kinetic energy is neglected).


\section{Numerical solution}
\label{sec:num}
There are numerous published methods for the numerical solution of Stefan problems, such as finite difference, finite element, level set, enthaply, integral and heat balance methods \cite{Caldwell04,Javierre06,Mitchell09, Mitchell14}. Our goal here is to demonstrate the form of solution and hence we will simply follow the finite difference scheme of  \cite{Font15}.

For the spherical problem, the change of variable $\mathrm{u}(r,t) = r T_l(r,t)$ and $\mathrm{v}(r,t) = r T_s(r,t)$ is made, which effectively transforms the problem geometry to Cartesian coordinates. The heat equations \eqref{NDheat} become
\subeq{
\begin{align}
\label{solid-numerics-1}
\rho c   \frac{\partial \mathrm{v}}{\partial t}  &= k\frac{\partial^2 \mathrm{v}}{\partial r^2}, &0 < \;\; &r < R(t),\\
\label{liquid-numerics-1}
\frac{\partial \mathrm{u}}{\partial t}  + v_l(r,t) \left(\frac{\partial \mathrm{u}}{\partial r} - \frac{\mathrm{u}}{r} \right) &= \frac{\partial^2 \mathrm{u}}{\partial r^2}, &R(t) < \;\; &r < R_b(t).
\end{align}
}
The boundary conditions transform to
\subeq{
\begin{align}
\mathrm{v}(0,t) &= 0, \\
\mathrm{v}(R,t) &= \mathrm{u}(R,t) = -\Gamma, \\
\frac{\partial \mathrm{u}}{\partial r}\bigg|_{r=R_b} &= Nu [R_b - \mathrm{u}(R_b,t)] +\frac{\mathrm{u}(R_b,t)}{R_b} ~,
\end{align}
}
and the Stefan condition \eqref{NDStef} is
\begin{align}
\rho \beta R \left[ 1- \frac{\gamma \Gamma}{R} -\frac{\delta}{2} \left( \frac{\ud R}{\ud t} \right)^2\right]\nd{R}{t} = k \frac{\partial \mathrm{v}}{\partial r}\bigg|_{r=R} - \frac{\partial \mathrm{u}}{\partial r}\bigg|_{r=R} + \frac{\Gamma (k-1)}{R}.
\end{align}
We now immobilise the free boundaries by introducing space-like variables defined as $\zeta = r / R(t)$ and $\eta = (r - R(t)) / (R_b(t) - R(t))$ in the heat equations for the solid and liquid, \eqref{solid-numerics-1} and \eqref{liquid-numerics-1}, respectively, which leads to
\subeq{
\label{numerics-2}
\begin{align}
\label{solid-numerics-2}
&\rho c\left(\frac{\partial \mathrm{v}}{\partial t} - \frac{\zeta}{R}\nd{R}{t}\pad{\mathrm{v}}{\zeta} \right) = \frac{k}{R^2} \frac{\partial^2 \mathrm{v}}{\partial \zeta^2}, \\
\label{liquid-numerics-2}
&\pad{\mathrm{u}}{t} - \frac{1}{R_b-R}\left[(1-\eta)\nd{R}{t} + \eta \nd{R_b}{t}\right]
\pad{\mathrm{u}}{\eta}
&+ \frac{v_l(\eta,t)}{R_b - R} \left[\pad{\mathrm{u}}{\eta} - \frac{(R_b - R)\mathrm{u}}{(R_b - R)\eta + R}\right] = \frac{1}{(R_b - R)^2}\frac{\partial^2 \mathrm{u}}{\partial \eta^2}.
\end{align}
}
These equations hold in fixed domains given by $0 < \zeta < 1$ and $0 < \eta < 1$.  Furthermore, the liquid velocity in the new coordinate system is
\begin{align}
v_l(\eta,t) = -(\rho - 1)\frac{R^2}{[(R_b - R)\eta + R]^2}\nd{R}{t}.
\end{align}
The boundary conditions become
\begin{align}
\label{bc-numerics}
\mathrm{v}(0,t) &= 0, \\
\mathrm{v}(1,t) &= \mathrm{u}(0,t) = -\Gamma, \\
\frac{1}{R_b - R}\frac{\partial \mathrm{u}}{\partial \eta}\bigg|_{\eta=1} &= Nu [R_b - \mathrm{u}(1,t)] +\frac{\mathrm{u}(1,t)}{R_b} ~ ,
\end{align}
and the Stefan condition is
\begin{equation}
\label{stefan-numerics}
\rho \beta R \left[1 - \frac{\gamma \Gamma}{R} -\frac{\delta}{2}\left( \frac{\ud R}{\ud t} \right)^2\right]\nd{R}{t} = \frac{k}{R} \frac{\partial \mathrm{v}}{\partial \zeta}\bigg|_{\zeta = 1} - \frac{1}{R_b - R}\frac{\partial \mathrm{u}}{\partial \eta}\bigg|_{\eta=0} + \frac{\Gamma (k-1)}{R}.
\end{equation}

Numerically solving the model requires an initial condition for the liquid temperature.  This can be obtained from a small-time analysis of the governing equations, as detailed in Appendix \ref{sec:small_time}. Following the analysis of  Appendix \ref{sec:small_time} we impose at some time $t_* \ll 1$, the following initial conditions:
\subeq{
\label{ic-numerics}
\begin{align}
\mathrm{v}(\zeta, t_*) &= -\Gamma R \zeta, \\
\mathrm{u}(\eta, t_*) &= -\left\{\Gamma + \Gamma(1 - R) -  Nu(1 + \Gamma)(R_b - R) \eta\right\}  \left[(R_b - R)\eta + R\right], \\
R(t_*) &= 1 - C t_*, \\
R_b(t_*) &= 1 + (\rho - 1)C t_*,
\end{align}
}
where $C$ is the solution of the cubic equation  \eqref{eqn:A}.

Spatial derivatives in the heat equations \eqref{numerics-2} are discretised using standard second-order centered finite differences.  To avoid the use of ghost points, the first-order space derivatives with respect to $\eta$ and $\zeta$ in the Stefan condition \eqref{stefan-numerics} are approximated by second-order forwards and backwards difference formulae, respectively. The spatial coordinates $\zeta$ and $\eta$ are discretised into a grid of $I$ nodes.  Semi-implicit time stepping of the heat equations is performed, whereby all of the terms involving $R$, $R_b$, and their derivatives are treated explicitly, and all terms involving the temperature are treated implicitly.  Therefore, at each time step, we solve an $2I \times 2I$ linear system, given by the discretisation of \eqref{numerics-2}, to determine the new temperature profiles.  We then substitute the temperatures into the Stefan condition \eqref{stefan-numerics} and solve the resulting cubic equation for $\ud R / \ud t$ to obtain the radius $R$ at the next time step using an explicit update. Further details may be found in \cite{Font15}.

\subsection{Transformed one-phase models}

The governing equation for the liquid is given by \eqref{liquid-numerics-2} with boundary and initial conditions \eqref{bc-numerics} and \eqref{ic-numerics}, respectively. The problem is closed via the Stefan condition
\begin{equation}
\label{stefan-numerics-largek}
\rho \beta R \left[1 - \frac{\gamma \Gamma}{R} - \frac{\delta}{2} \left(\frac{\ud R}{\ud t}\right)^2   - \frac{c \Gamma}{3R} \right]\nd{R}{t} = - \frac{1}{R_b-R} \frac{\partial \mathrm{u}}{\partial \eta}\bigg|_{\eta=0} - \frac{\Gamma}{R} ~,
\end{equation}
which is accurate to $\ord{1/k^2}$. If the effect of the solid is to be neglected completely then the final term in the square brackets disappears.

\section{Results and discussion}

The model proposed in Section \ref{sec:sphere} as well as its reductions will now be used to study the melting of spherical gold nanoparticles. Gold has been chosen because it is one of the most commonly used nanomaterials and, as can be seen from Table \ref{tab:tin}, it has a relatively large contrast between the density and thermal conductivity of each phase. The governing equations will be solved using the numerical scheme described in Section \ref{sec:num} and the parameter values provided in Table \ref{tab:tin}. This includes the maximum value of heat transfer coefficient which although unrealistically high leads to results in line with those previously published in the literature. We first focus on the role of density variation during melting. Then, the validity of the  large Stefan number reduction is explored. Finally, we compare the two one-phase models ($k \gg 1$ and $k \to \infty$) to the full two-phase model.

\subsection{Impact of density variation during melting}

Font \etal\cite{Font15} demonstrated that including density change in the model for the melting of a gold nanoparticle will significantly slow the melting process. This is primarily due to the addition of kinetic energy to the Stefan condition. Melting times more than double those of the constant density model  were predicted. This study applied a fixed-temperature boundary condition, which results in an initial infinite melt rate, so it not so surprising that kinetic energy played an important role.  Ribera and Myers \cite{Ribera2016} subsequently studied the melting of tin subject to a  Newton condition \eqref{bc:newton}. In this case the difference between variable-density and constant-density models was only a few percent. This change may be attributed to two factors, firstly the Newton condition does not lead to an infinite melt rate, secondly tin has a much  smaller density contrast than gold. However, it clearly demonstrated that the effect of density change should not be as severe as suggested by Font \emph{et al.} \cite{Font15}. Unfortunately, as discussed in \S \ref{IESec} the models used in both \cite{Font15, Ribera2016} were derived from that of the textbook \cite{Alex} which incorrectly represents the kinetic energy. As shown by equation \eqref{FinStef2} the term $(\rho_s/2)(1 - \rho_s / \rho_l)^2 R_t^3$ used in the previous studies should be replaced with $(\rho_s/2)[1-(\rho_s / \rho_l)^2 ] R_t^3$. By comparing the coefficients of $R_t^3$ in these terms (with the parameter values of Table \ref{tab:tin}) we find that kinetic energy in the present model is roughly 18 times stronger in the case of gold, and 70 times stronger in the case of tin, indicating that the role of density variation needs to be re-assessed.

\begin{figure}
  \centering
  \subfigure[$R_0 = 10$~nm]{\includegraphics[width=0.49\textwidth]{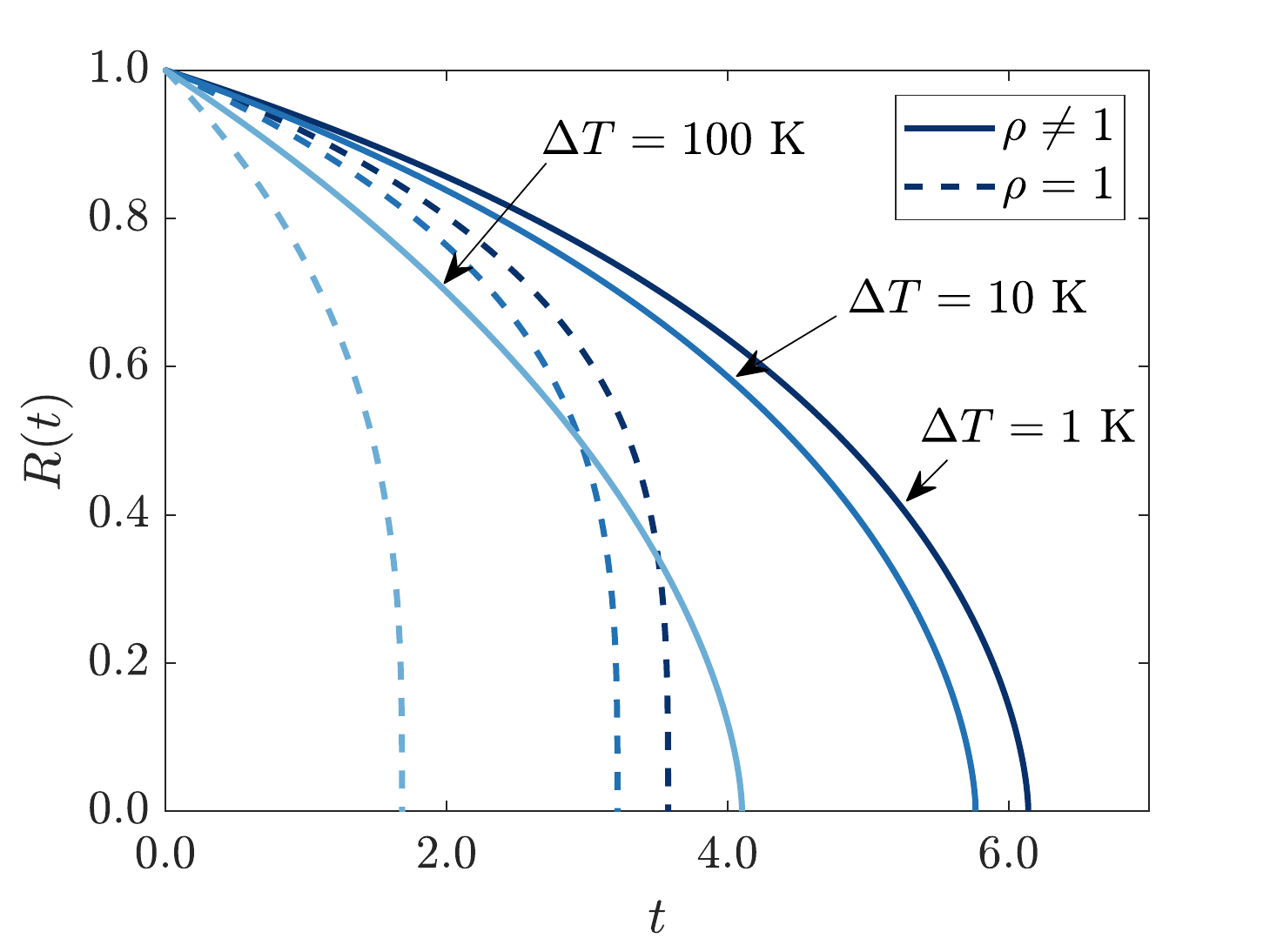}}
  \subfigure[$R_0 = 100$~nm]{\includegraphics[width=0.49\textwidth]{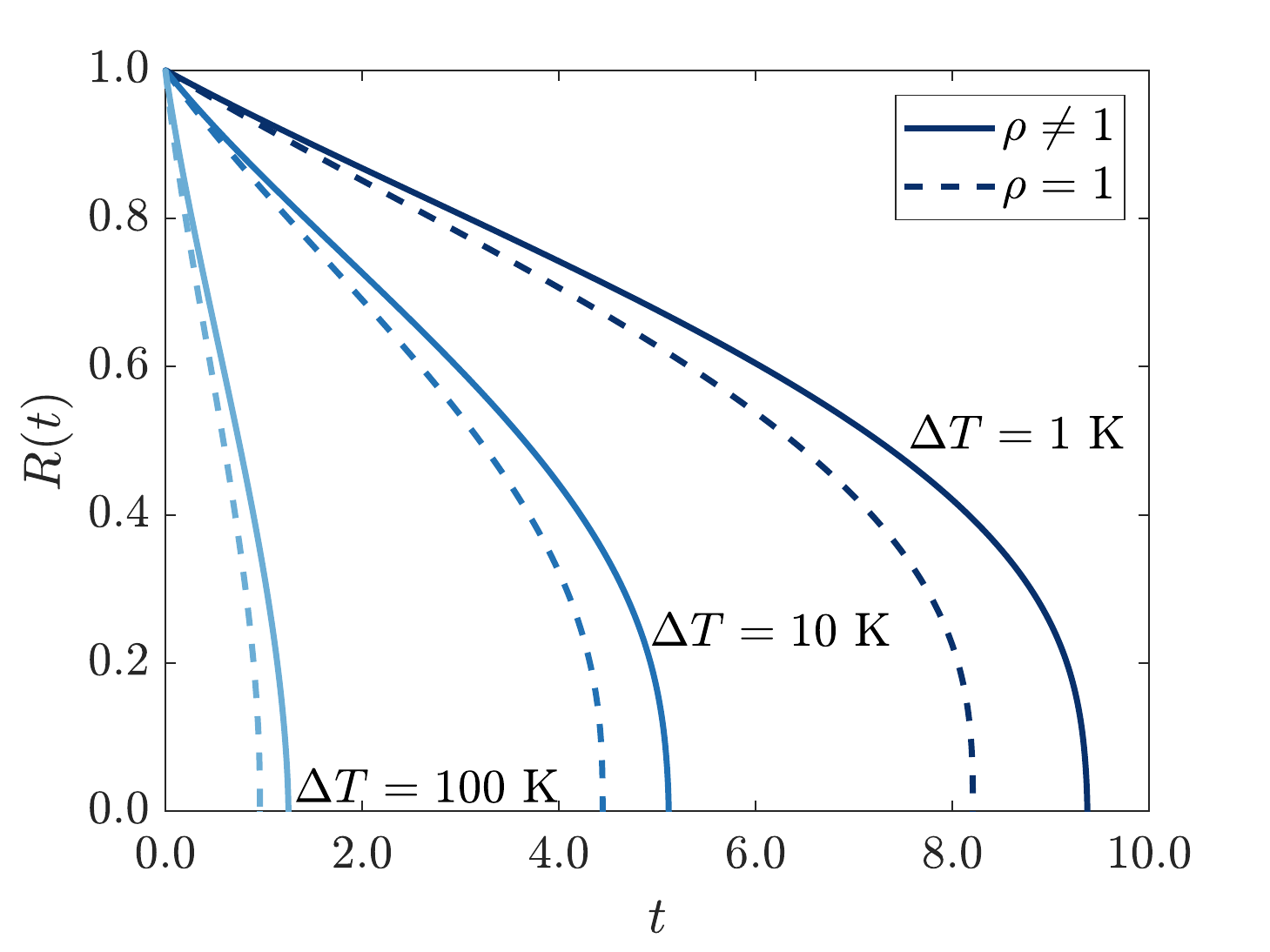}}
  \caption{Numerical simulations of nanoparticle melting that do (solid lines) and do not (dashed  lines) consider density variation between the solid and liquid. The parameter values are based on gold and can be found in Table \ref{tab:tin}; the model can be found in Section \ref{sec:sphere}.}
  \label{fig:compare_rho}
\end{figure}

Figure \ref{fig:compare_rho} displays the numerical solution for the variation of the radius with time with (solid lines) and without (dashed lines) density variation in the model. Initial radii of 10~nm and 100~nm were considered for various imposed temperature differences $\Delta T$. As stated, with density variation kinetic energy removes some of the energy available for melting which then acts to slow down the process. From the figure it is clear that for the  10~nm particle the differences in predictions are large, the melting time increases by 72\%, 80\%, and 140\% for temperature differences of $\Delta T = 1$~K, $10$~K, and $100$~K, respectively. With the 100~nm particles the effect is not so dramatic with increases of 14\%, 15\%, and 30\% for $\Delta T = 1$~K, $10$~K, and $100$~K.

Increasing the initial radius, up to 10~$\mu$m, the melting time settles at approximately 15\% higher than the fixed-density model for all tested values of $\Delta T$. From this we see, in line with the conclusions of \cite{Font15}, that density plays an important role in the melting process well beyond the nanoscale and so should not be neglected in any practical study. Decreasing the heat transfer coefficient leads to reductions in the discrepancy: if we take $h = 10^{-5}\,h_\text{max}$ then the difference in melt time is around 8\% for large particles. Note, the effect of kinetic energy is reflected in the size of $\delta \propto 1/R^2$. For a 10~$\mu$m gold nanoparticle $\delta \approx 10^{-6}$, which is obviously negligible. With a small heat transfer coefficient there is no initial high velocity, so the differences cannot be due to kinetic energy. Density change also affects the position of the outer boundary, consequently we conclude that the difference in melt time for large particles must be due primarily to the fact that the fixed-density model gives errors in the boundary position, which in turn affects the domain over which the external heat must travel.

\subsection{Validation of reduced models for large Stefan number}

The reduction of the model in the limit of large Stefan number was described in Section \ref{sec:large_beta}. This led to a single differential equation for the position of the melt front, equation \eqref{ODE_largebeta}. It was then shown that with the present (maximal) value of heat transfer coefficient, the initial size of the kinetic energy term is $K \approx 0.15$. Thus, neglecting kinetic energy may lead to errors on the order of 15\% from the very beginning of the melt process. Consequently, a second differential equation which accounts for kinetic energy, given by \eqref{ODE_largebeta_ke}, was presented. As the Nusselt number decreases this equation reduces to the standard large Stefan number form.

\begin{figure}
  \centering
  \subfigure[$R_0 = 10$~nm]{\includegraphics[width=0.49\textwidth]{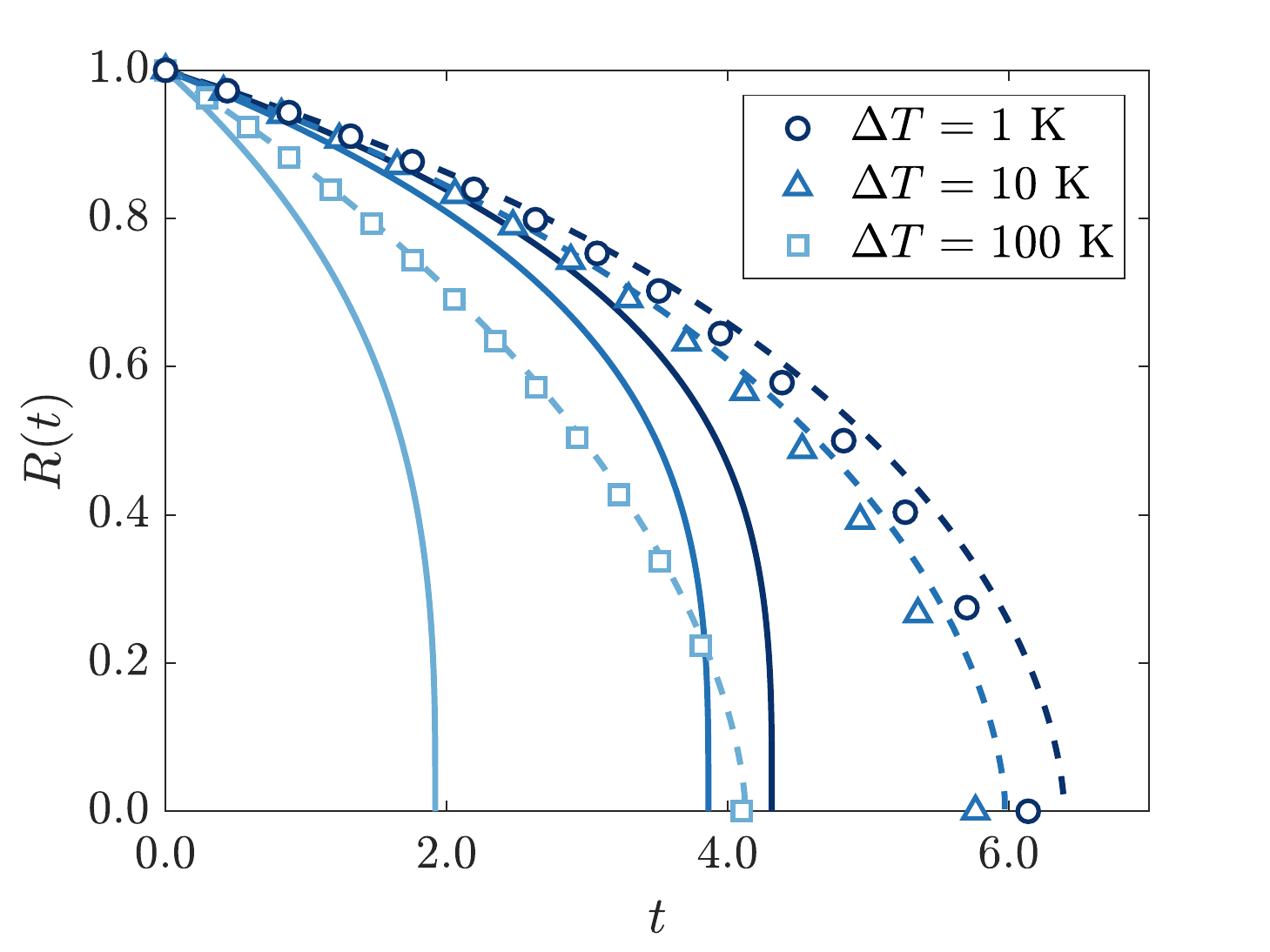}}
  \subfigure[$R_0 = 100$~nm]{\includegraphics[width=0.49\textwidth]{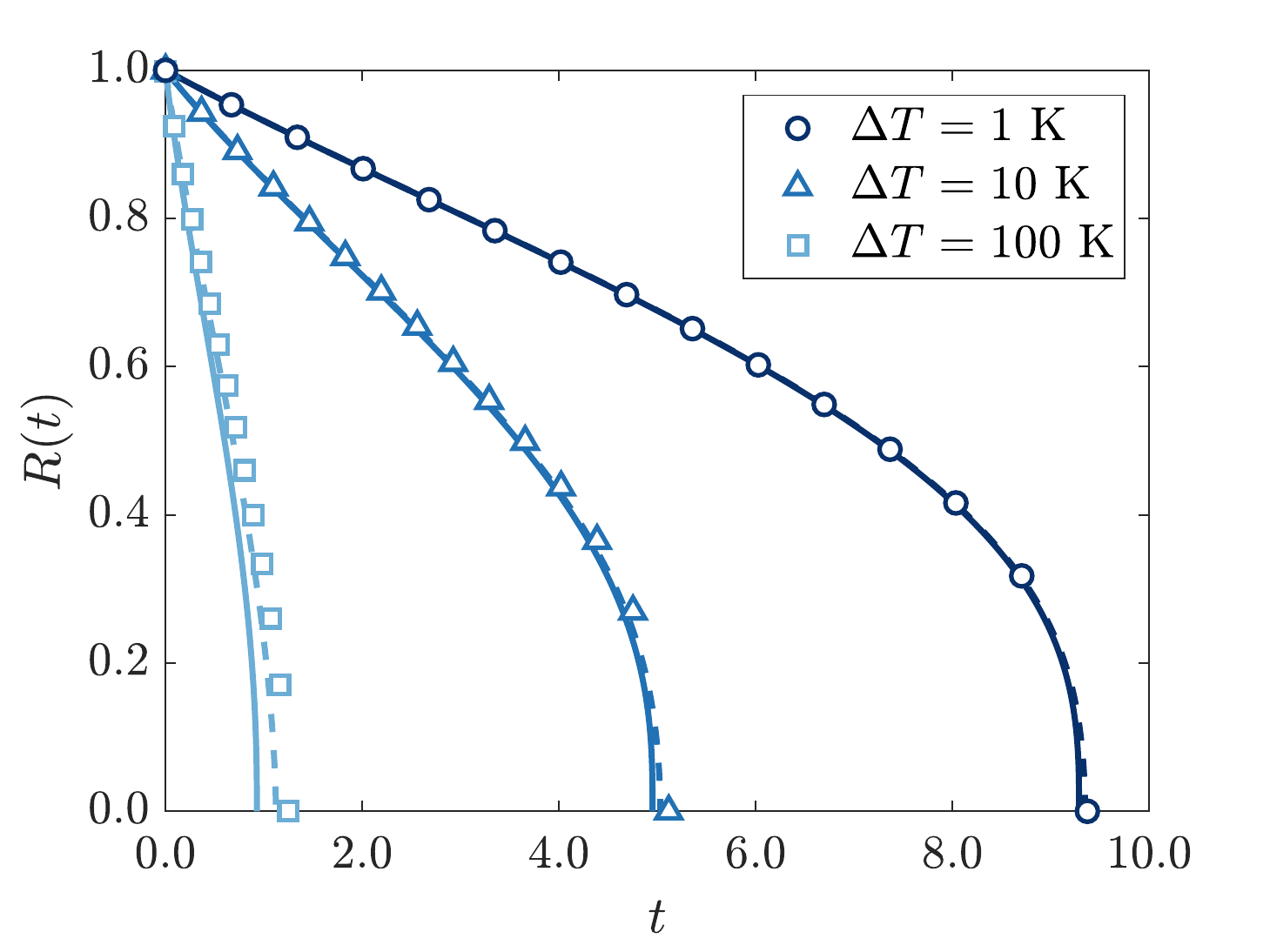}}
  \caption{Numerical simulations of nanoparticle melting using the full model and reduced models that are valid in the limit of large Stefan number. Symbols, solid lines, and dashed lines denote solutions of the full model, reduced model without kinetic energy \eqref{ODE_largebeta}, and reduced model with kinetic energy \eqref{ODE_largebeta_ke}.  Parameter values correspond to gold. The Stefan numbers for $\Delta T = $ 1~K, 10~K, and 100~K are $\beta = 390$, $39$, and $3.9$, respectively.}
  \label{fig:large_beta}
\end{figure}

In Figure \ref{fig:large_beta} (a)--(b), we compare the numerical solution of the full model with the predictions of the large Stefan number reduction, with and without kinetic energy: symbols denote the numerical solution, solid lines the standard reduction, and dashed lines the reduction including the kinetic energy term. As before, we consider gold nanoparticles with initial radii of 10~nm and 100~nm. The $\Delta T$ values of 1~K, 10~K, and 100~K correspond to Stefan numbers $\beta = 390$, $39$, and $3.9$, respectively.  With such high values of $\beta = 390, 39$ we would expect almost exact agreement between all solutions, yet in the case of a 10~nm particle, the deviation between the full numerical and simple reduction  is substantial, with errors in the melting times of 30\%, 33\%, and 53\% for $\Delta T =$ 1~K, 10~K, and 100~K.  Retaining kinetic energy results in much closer agreement with errors in the melting time of 4.1\%, 3.6\%, and 0.46\%. With a 100~nm particle the value of $K$ is significantly lower (due to the reduction in $\delta$ by a factor of 100) and then both large Stefan reductions show excellent agreement with the numerical solution.

The importance of density change can be observed by comparing Figures \ref{fig:compare_rho}~(b) and \ref{fig:large_beta}~(b). Neglecting the density change leads to melting-time errors between 14 and 30\% in Figure \ref{fig:compare_rho}~(b), whereas in Figure \ref{fig:large_beta}~(b) the results are almost identical. Figure \ref{fig:large_beta}~(b) demonstrates that for larger particles kinetic energy does not play such an important role and thus confirms the previous conclusion that errors, for sufficiently large particles, are a consequence of the fact that the constant-density model does not capture the growth of the outer boundary at $r = R_b(t)$. Practically this means that for sufficiently large material samples we could use an equation of the form \eqref{ODE_largebeta}, i.e. neglecting kinetic energy, but retaining the density difference in the term representing the motion of the outer boundary.

\subsection{Comparison of one- and two-phase models}

\begin{figure}
  \centering
  {\includegraphics[width=0.49\textwidth]{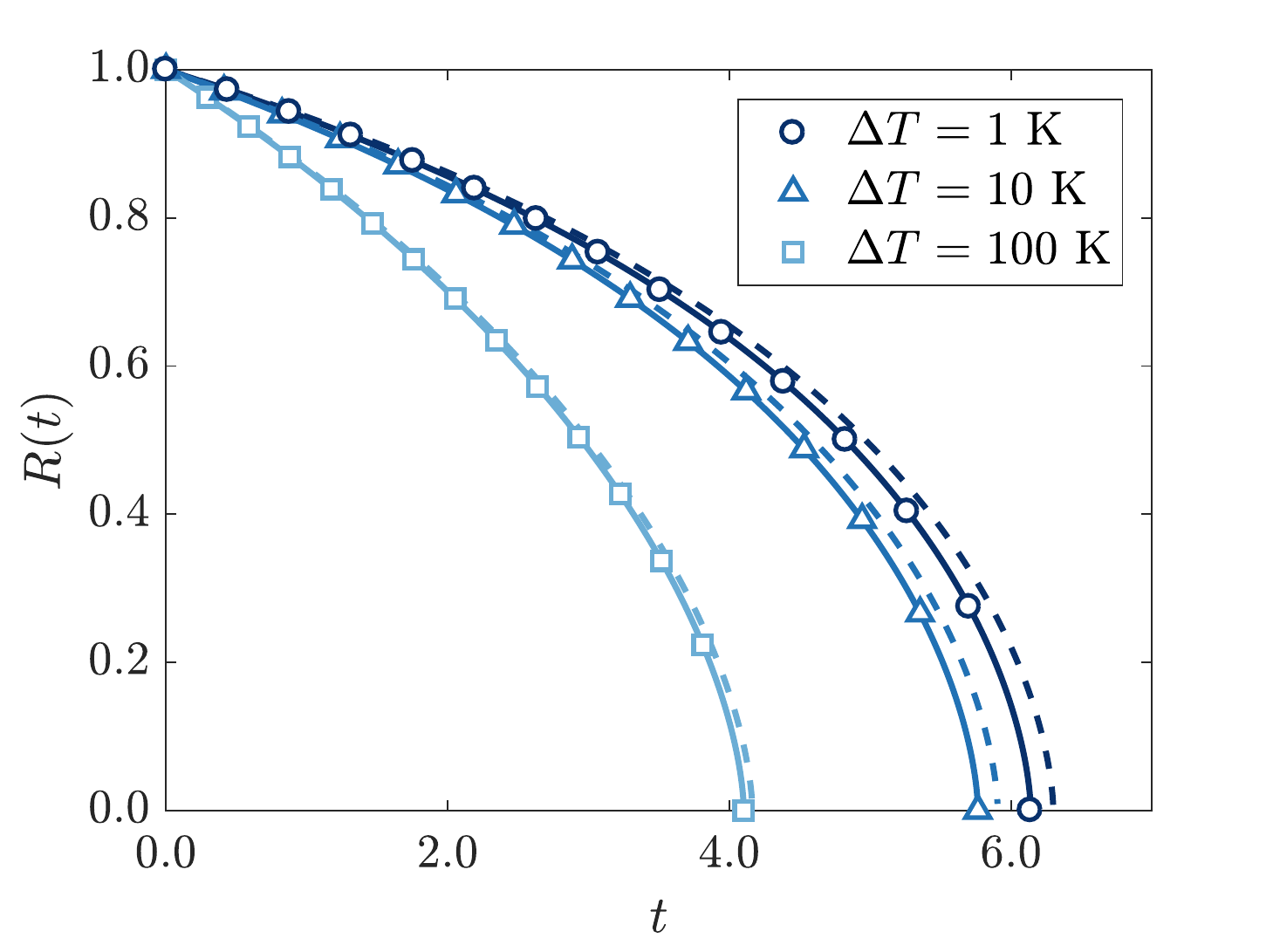}}
  \caption{Numerical simulations of nanoparticle melting using the full two-phase model (symbols), the one-phase model valid for large $k$ (solid lines), and the one-phase model that neglects the solid altogether (dashed lines) and assumes $T_s = T_I(R(t))$. Parameter values correspond to gold, which has $k \simeq 3.0$ and an initial radius of 10~nm was used.}
  \label{fig:large_k}
\end{figure}

Two approaches for performing a one-phase reduction of the full two-phase model are discussed in Section \ref{sec:one_phase}.  The first approach is based on systematically exploiting the large ratio of solid to liquid thermal conductivity, $k \gg 1$, and led to the modified Stefan condition given by \eqref{1PStef}. The second approach is based on simply neglecting the solid phase altogether and setting $T_s = T_I(R(t))$, which is technically only valid in the limit $k \to \infty$. The accuracy of the reduced one-phase models is now assessed through comparison with the full two-phase model. Figure \ref{fig:large_k} shows numerical simulations of the one- and two-phase models for various values of $\Delta T$ for a fixed value of $R_0 = 10$~nm. The `asymptotic' one-phase model based on $k \gg 1$, shown as solid lines, is exceptionally accurate; in all cases the solutions perfectly coincide with those of the two-phase model (shown as symbols). The high degree of accuracy is remarkable given that the ratio of thermal conductivity for gold is $k \sim 3.0$, which is not particularly large. Even the one-phase model that treats the solid temperature as uniform (\emph{i.e.} assumes $k \to \infty$, shown as dashed lines) is accurate, and overestimates the melting time by only 2\%. This overestimate occurs because the conduction of heat from the bulk of the solid to the interface, which has a lower temperture than the bulk due to size effects, is neglected from this version of the one-phase model. If the initial radius of the nanoparticle is increased to 100~nm, then both one-phase models are in perfect agreement with the two-phase model.

\section{Non-Fourier heat transfer}\label{NonFSec}

The results presented so far have been in non-dimensional form and so hide the fact that melting times for the 10~nm particle are on the order of picoseconds. This is comparable to the thermal relaxation time for many materials. Practically this means that Fourier's law may no longer accurately describe the transfer of heat that occurs with phase change at very small length (and also time) scales.
The breakdown of Fourier's law, at both small  length and time scales, has been predicted theoretically, via molecular dynamics, and observed experimentally, see \cite{Guo18,Yang10}. Consequently we will now briefly mention alternative heat conduction models.

There are many different models that aim to correctly describe heat transfer in situations where Fourier's law is invalid, they can be categorized into micro-, meso- and macroscopic approaches. Micro- and mesoscopic models focus on each of the heat carriers or on their distribution, whereas the macroscopic models aim to describe the system in terms of quantities such as the temperature or the heat flux. The computational cost of approaches belonging to the first two categories, such as molecular dynamics \cite{Yang10}, the Boltzman transport equation (BTE) \cite{Boltzmann1872} or the equation of phonon radiative transfer (EPRT) \cite{Majumdar1991} is high and therefore only very small systems are usually considered.
An approach that blends well with the PDE formulation used so far is a macroscopic model of the form
\subeq{
\bea
\tau_R \pad{\mathbf{q}}{t} + \mathbf{q} =  - k \nabla T + \ell^2 (\nabla ^2 \mathbf{q} + 2 \nabla \nabla \cdot \mathbf{q}) ~ .
\lb{eqn:GK}
\eea
}
This is called the Guyer--Krumhansl or GK equation.
The first new parameter introduced here is $\tau_R$ which represents the thermal resistive relaxation time, i.e., the mean time between collisions among the heat carriers (phonons) which do not conserve their momentum; the second parameter $\ell$ is the phonon mean free path, i.e., the mean distance that a heat carrier travels between collisions. The first term in \eqref{eqn:GK} is said to add memory to the system and is dominant on time scales comparable to $\tau_R$, the second-order derivatives of the flux describe non-local effects that dominate heat transport in  systems with length scales comparable to $\ell$. In the limit $\ell \ra 0$ a simpler equation, the Maxwell--Cattaneo law, is retrieved \cite{Cattaneo1958,Vernotte1958}.  The Maxwell--Cattaneo equation gives rise to the hyperbolic or relativistic heat equation, which describes heat propagation with finite speed. Although the Guyer--Krumhansl equation was originally derived for situations with extremely low temperatures, recent studies have shown that it may be accurate at much higher temperatures \cite{Guo2018}. It has been used to successfully predict the size dependence of the thermal conductivity in nanowires for temperatures ranging from 1 to 350 K \cite{Calvo2018,Calvo2018b}, a feature that cannot be captured using Fourier's law. The effect of using the Guyer-Krumhansl equation on phase-change models has been studied in detail by Hennessy et al. \cite{Hennessy2018b}. The Maxwell--Cattaneo equation is used in the framework of melting nanoparticles in \cite{Hennessy2018}.

The use of a non-Fourier conduction law in a phase-change model can result in a temperature jump across the interface formed between the two phases \cite{Greenberg1987, Sobolev1991, Sobolev1996}. When Fourier's law is replaced by the Maxwell--Cattaneo equation, this temperature jump is needed to ensure that the speed of the interface is less than the finite speed at which thermal energy can be delivered to it. For example in the one-dimensional experimental setup described in \cite{Hennessy2018b} the temperature jump is given by
\begin{align}
  T_l - T_s = \frac{L_m}{c(\mathcal{V}^2 - s_t^2)}\left(s_t^2 - \frac{1}{\rho L_m \tau_R}(Q_l-Q_s)\right), \quad x = s(t),
\end{align}
where $\mathcal{V} = (\alpha / \tau_R)^{1/2}$ is the speed of hyperbolic heat transport and $Q = -3 \ell (\partial q/\partial x)$ is the flux of the flux \cite{Jou2010}.

\begin{figure}[h!]
	\centering
	\includegraphics[width=.49\textwidth]{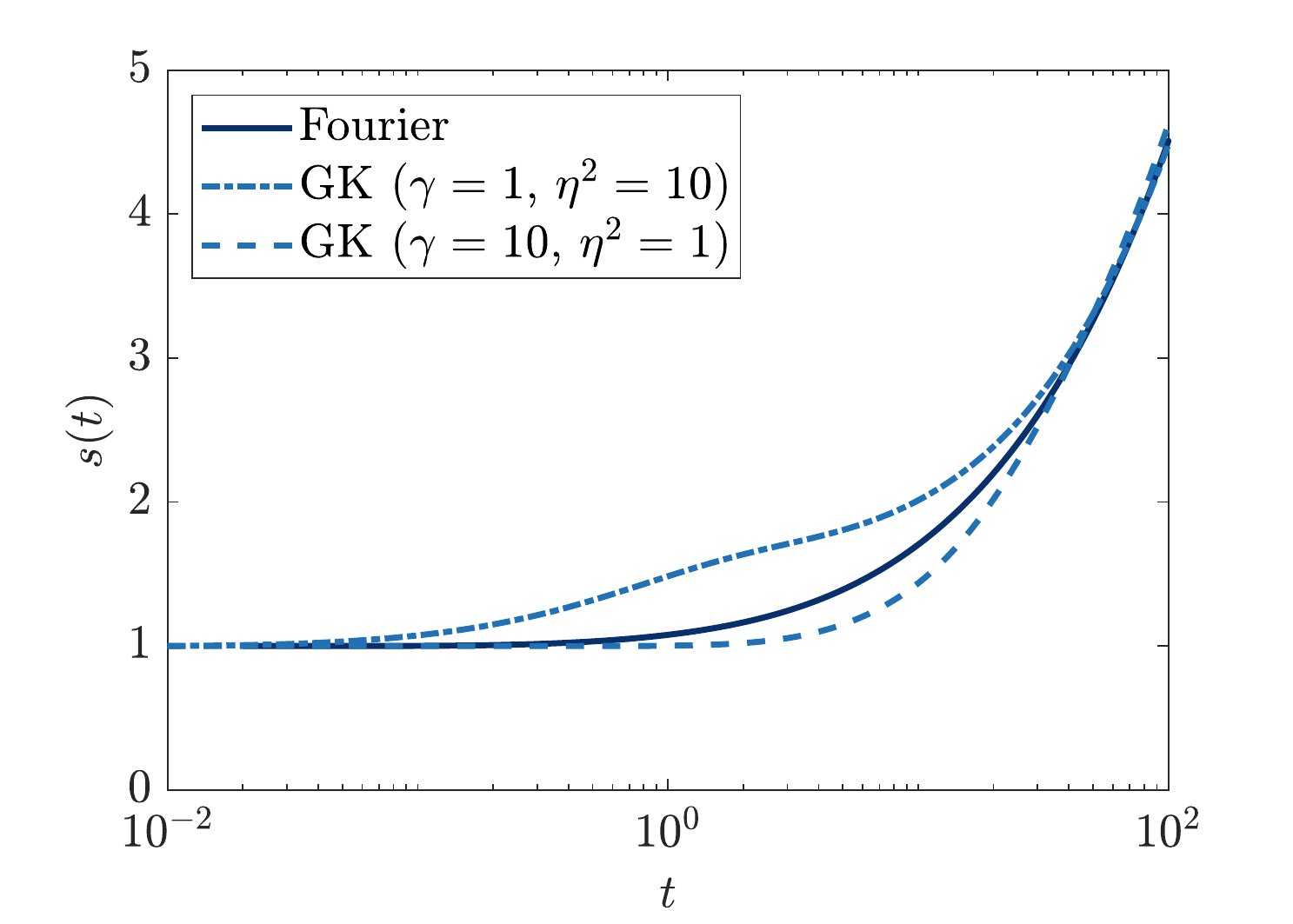}
	\caption{Evolution of the solid-liquid interface in a semi-infinite liquid bath which starts solidifying from a seed crystal due to a cold environment, as predicted by Fourier's law or by the Guyer-Krumhansl equation with different values of the non-dimensional parameters $\gamma$ and $\eta^2$. The Stefan number has been set to $\beta=10$.}\label{fig:NonFourier}
\end{figure}

In Fig. \ref{fig:NonFourier} we present typical solutions for the position of a solidification front predicted via the Guyer--Krumhansl equation using the finite-difference method described in \cite{Calvo2019}: we refer to this paper for more details on the scheme. In this case a one-phase situation is assumed, allowing us to write
\begin{align}
\pad{T_s}{t} + \pad{q_s}{x} & = 0 ~,\\
\gamma \pad{q_s}{t} + q_s &= - \pad{T_s}{x} + \eta^2 \padd{q_s}{x} ~ .
\end{align}
where $\gamma$ and $\eta$ are the non-dimensional relaxation time and the mean free path. For the simulation
the boundary temperature was fixed $T(0,t)=-1$ and the temperature at the interface constant $T(s,t)=0$, initially $s = 1$ with $T=q=0$.
The position of the solid-liquid interface is plotted for typical values of  $\gamma$ and  $\eta$ as well as the  classical case ($\gamma=\eta=0$). When the relaxation time is high, $\gamma=10$, it takes longer for heat to be transmitted from the boundary to the front and so solidification is slower than in the classical case. If non-local effects were neglected $\eta=0$ (which holds for the Maxwell--Catteneo model) this would be true for any relaxation time $\gamma > 0$. However, from the curve with $\eta^2 = 10$ we see that the diffusion of heat flux can make the solidification occur more rapidly, despite the fact $\gamma > 0$. The figure makes it clear that non-Fourier effects can play a significant role at small length and time scales, however for sufficiently large times (and hence sufficiently large length-scale $s(t)$) we expect non-Fourier effects to disappear, this is apparent as all solutions coincide for large times.

\section{Conclusions}

The governing equations describing the phase change of a material with distinct thermophysical properties in each phase were derived from conservation laws. This is a natural approach for a two-phase mixture, with a finite region over which properties change. As the width of this region approaches zero the sharp-interface limit, used in many Stefan problems, is retrieved. The one drawback to this approach is that being the limit of a mixture model there is no mechanism to account for surface tension. This could be dealt with formally by introducing an appropriate Korteweg stress tensor, here we simply added a source term at the interface. From the conservation laws the standard heat equations are obtained. Applying the Rankine--Hugoniot condition to the conservation laws leads to relations between the material and interface velocities, the pressure and energy jump. The latter expression provides the Stefan condition.

An important point to note is that the obtained Stefan condition differs to those found in the literature. Various expressions, with minor differences to that derived here, have been provided for the effective latent heat and the contribution of the difference between interface and bulk melt temperature. A particular error concerns the kinetic energy contribution which has subsequently propagated into subsequent studies. The Stefan condition derived here clarifies the various energetic contributions and is written in such a way that it may be readily incorporated into non-Fourier based studies as well.

To illustrate the use of the governing equations, the system was first applied to the standard problem of solidification of a semi-infinite bar and subsequently to the popular problem of a melting spherical nanoparticle. The nanoparticle case is particularly useful, since it permits the study of size dependent parameters in a simple one-dimensional setting.
Using this vehicle, we were able to determine the significant effect of density change on melting. At small length-scales it is very large and the effect continues into the macroscale with errors of the order 10\%  observed. It was shown that for small particles kinetic energy plays an important role, while at larger scales it is primarily the error in the position of the outer boundary that leads to incorrect melting times. Overall, the difference between the fixed- and changing-density model depends on the relative density of the two phases, the heat transfer at the boundary, and the initial particle size.

Although the examples were primarily focused on the case of spherically symmetric nanoparticle melting the formulation is such that it can be readily applied to other practical geometries where size-dependent material properties play an important role, such as the case of cylindrical nanowire melting \cite{Florio2016}. The evolution of micro- and macro-scopic systems can also be influenced by size-dependent material properties, a well-known example of which is the stabilisation of solid dendritic structures forming in a supercooled melt \cite{Davis2001} by the change in melt temperature with curvature.

Finally we briefly discussed the problem of heat flow and phase change in situations where Fourier's law no longer holds. This is becoming increasingly important as advances are made in nanotechnology. This field is currently wide open, with many opportunities for mathematical advancement into exciting new and practically interesting problems.

\section*{Acknowledgements}

This project has received funding from the European Union's Horizon 2020 research and innovation
programme under grant agreement No 707658. MC acknowledges that the research leading to these results
has received funding from `la Caixa' Foundation. TM acknowledges financial support from the Ministerio
de Ciencia e  Innovaci\'{o}n Grant No. MTM2017-82317-P.
The authors have been partially funded by the CERCA Programme of the Generalitat de Catalunya.

\begin{appendix}

\section{Small-time solution}
\label{sec:small_time}

We may focus on the small-time behaviour by
introducing an artificial small parameter $\epsilon \ll 1$ and then re-scaling time $t = \epsilon \bar{t}$. The radius $R$ is obviously close to its initial value, $R(t) = 1 - \epsilon \bar{R}(\bar{t})$.  An expansion of \eqref{eqn:nd_Rb} in terms of $\epsilon$ shows that $R_b = 1+ \epsilon (\rho - 1) \bar{R} + O(\epsilon^2)$.   The Gibbs--Thomson relation can be
expanded as $T_m = -\Gamma / R = -\Gamma - \epsilon \Gamma \bar{R} + O(\epsilon^2)$.  Upon writing $T_s(r,t) = \bar{T}_s(r,\bar{t})$, the leading-order problem for the temperature of the solid becomes
$\partial \bar{T}_s / \partial \bar{t} = 0$, with boundary and initial conditions given by
$\partial \bar{T}_s / \partial r = 0$ at $r = 0$, $\bar{T}_s(1,\bar{t}) = -\Gamma$, and $\bar{T}_s(r,0) = -\Gamma$.
The appropriate solution is then $\bar{T}_s(r,t) = -\Gamma$ and is constant.

Since the liquid layer is initially very thin we must scale the radial co-ordinate in accordance with the variation of $R$, hence $r = 1 - \epsilon \bar{r}$. The liquid temperature is written as $T_l(r,t) = -\Gamma + \epsilon \bar{T}_l(\bar{r},\bar{t})$,
which ensures that there is a balance of terms in the Newton condition.  The leading-order problem for the liquid temperature is then
\begin{align}
\padd{\bar{T}_l}{\bar{r}} = 0,
\end{align}
subject to
\subeq{
\begin{align}
-\left.\pad{\bar{T}_l}{\bar{r}}\right|_{\bar{r}=-\bar{R}_b(\bar{t})}  = Nu(1 + \Gamma) ~, \qquad
\bar{T}_l(\bar{R}(\bar{t}),\bar{t}) = -\Gamma \bar{R}(\bar{t}) ~.
\end{align}
}
This has solution
\begin{align}
\bar{T}_l(\bar{r},\bar{t}) = - \Gamma \bar{R}(\bar{t}) -  Nu(1 + \Gamma)\left[\bar{r} - \bar{R}(\bar{t})\right].
\end{align}

Since the leading order solid temperature is constant for small times  it does not enter the Stefan condition and the melting is driven by the conduction of heat from the exterior through the liquid.
The leading-order Stefan condition can be written as
\begin{align}
\rho \beta \left[1-\gamma \Gamma - \frac{\delta}{2}\left(\nd{\bar{R}}{\bar{t}}\right)^2\right]\nd{\bar{R}}{\bar{t}} = Nu (1 + \Gamma) ~ .
\end{align}
At small times the solution typically takes a power law form $\bar{R} = C \bar{t}^{\alpha}$. Here we find $\alpha=1$ and
$C$ satisfies the cubic equation
\begin{align}
\rho \beta \left[1-\gamma \Gamma - \frac{\delta}{2}C^2\right]C = Nu (1 + \Gamma)\label{eqn:A} ~ .
\end{align}

The small-time solution for the temperature and positions of the interfaces, in terms of the original dimensionless variables, can be written as
\subeq{
\begin{align}
T_s(r,t) &= -\Gamma, \\
T_l(r,t) &= -\Gamma - \Gamma(1 - R(t)) - Nu (1 + \Gamma) (R(t) - r), \\
R(t) &= 1 - C t, \\
R_b(t) &= 1 + (\rho - 1) C t,
\end{align}
}
where $C$ satisfies \eqref{eqn:A}.

\end{appendix}

\bibliographystyle{abbrv}
\inputencoding{utf8}
\bibliography{NewStefBib}

\end{document}